\documentclass[table,twocolumn,mathlines]{aastex631}
\usepackage[utf8]{inputenc}
\usepackage{rotating}
\usepackage{color}
\usepackage{graphicx}
\usepackage{amsmath}
\usepackage{subfigure}
\usepackage{array}
\usepackage{comment}
\usepackage{soulutf8} 
\usepackage{xcolor}
\usepackage{ulem}  

\newcommand{\rmsub}[1]{\ensuremath{_{\mathrm{#1}}}}

\newcommand{\yr}{\textrm{yr}} 
\newcommand{\ps}{\ensuremath{\mathrm{s^{-1}}}} 
\newcommand{\km}{\textrm{km}} 
\newcommand{\pc}{\textrm{pc}} 
\newcommand{\kpc}{\textrm{kpc}} 
\newcommand{\dmu}{\ensuremath{\mathrm{pc\, cm^{-3}}}}
\newcommand{\Msun}{\ensuremath{\mathrm{M_{\Sun}}}} 

\DeclareUnicodeCharacter{02BC}{'}

\begin{document}

\shorttitle{The FAST Globular Cluster Pulsar Survey (GC FANS)}
\shortauthors{Lian, et al.}

\title{The FAST Globular Cluster Pulsar Survey (GC FANS)}

\author{Yujie Lian}
\affiliation{Institute for Frontiers in Astronomy and Astrophysics, Beijing Normal University, Beijing 102206, Peopleʼs Republic of China}
\affiliation{School of Physics and Astronomy, Beijing Normal University, Beijing 100875, Peopleʼs Republic of China}
\affiliation{Max-Planck-Institut f\"ur Radioastronomie, Auf dem H\"ugel 69, D-53121 Bonn, Germany}

\author{Zhichen Pan$^{\ast}$}
\affiliation{National Astronomical Observatories, Chinese Academy of Sciences, 20A Datun Road, Chaoyang District, Beijing, 100101, Peopleʼs Republic of China}
\affiliation{Guizhou Radio Astronomical Observatory, Guizhou University, Guiyang 550025, Peopleʼs Republic of China}
\affiliation{Key Laboratory of Radio Astronomy and Technology (Chinese Academy of Sciences), A20 Datun Road, Chaoyang District, Beijing 100101, Peopleʼs Republic of China}
\affiliation{College of Astronomy and Space Sciences, University of Chinese Academy of Sciences, Beijing 100049, Peopleʼs Republic of China}

\author{Haiyan Zhang$^{\dag}$}
\affiliation{National Astronomical Observatories, Chinese Academy of Sciences, 20A Datun Road, Chaoyang District, Beijing, 100101, Peopleʼs Republic of China}
\affiliation{Key Laboratory of Radio Astronomy and Technology (Chinese Academy of Sciences), A20 Datun Road, Chaoyang District, Beijing 100101, Peopleʼs Republic of China}
\affiliation{College of Astronomy and Space Sciences, University of Chinese Academy of Sciences, Beijing 100049, Peopleʼs Republic of China}

\author{Shuo Cao$^{\ddag}$} 
\affiliation{Institute for Frontiers in Astronomy and Astrophysics, Beijing Normal University, Beijing 102206, Peopleʼs Republic of China}
\affiliation{School of Physics and Astronomy, Beijing Normal University, Beijing 100875, Peopleʼs Republic of China}

\author{P. C. C. Freire$^{\S}$}
\affiliation{Max-Planck-Institut f\"ur Radioastronomie, Auf dem H\"ugel 69, D-53121 Bonn, Germany}

\author{Lei Qian}
\affiliation{National Astronomical Observatories, Chinese Academy of Sciences, 20A Datun Road, Chaoyang District, Beijing, 100101, Peopleʼs Republic of China}
\affiliation{Guizhou Radio Astronomical Observatory, Guizhou University, Guiyang 550025, Peopleʼs Republic of China}
\affiliation{Key Laboratory of Radio Astronomy and Technology (Chinese Academy of Sciences), A20 Datun Road, Chaoyang District, Beijing 100101, Peopleʼs Republic of China}
\affiliation{College of Astronomy and Space Sciences, University of Chinese Academy of Sciences, Beijing 100049, Peopleʼs Republic of China}

\author{Ralph~P.~Eatough}
\affiliation{National Astronomical Observatories, Chinese Academy of Sciences, 20A Datun Road, Chaoyang District, Beijing, 100101, Peopleʼs Republic of China}
\affiliation{Max-Planck-Institut f\"ur Radioastronomie, Auf dem H\"ugel 69, D-53121 Bonn, Germany}

\author{Lijing Shao}
\affiliation{Kavli Institute for Astronomy and Astrophysics, Peking University, Beijing 100871, Peopleʼs Republic of China}
\affiliation{National Astronomical Observatories, Chinese Academy of Sciences, 20A Datun Road, Chaoyang District, Beijing, 100101, Peopleʼs Republic of China}
\affiliation{Max-Planck-Institut f\"ur Radioastronomie, Auf dem H\"ugel 69, D-53121 Bonn, Germany}

\author{Scott M. Ransom}
\affiliation{National Radio Astronomy Observatory, Charlottesville, VA 22903, USA}

\author{Duncan R. Lorimer}
\affiliation{Department of Physics and Astronomy, West Virginia University, Morgantown, WV 26506-6315, USA}
\affiliation{Center for Gravitational Waves and Cosmology, West Virginia University, Chestnut Ridge Research Building, Morgantown, WV 26505, USA}

\author{Dejiang Yin}
\affiliation{College of Physics, Guizhou University, Guiyang 550025, Peopleʼs Republic of China}

\author{Yinfeng Dai}
\affiliation{College of Physics, Guizhou University, Guiyang 550025, Peopleʼs Republic of China}

\author[0000-0002-2953-7376]{Kuo Liu}
\affiliation{Shanghai Astronomical Observatory, Chinese Academy of Sciences, 80 Nandan Road, Shanghai 200030, Peopleʼs Republic of China}
\affiliation{Key Laboratory of Radio Astronomy and Technology (Chinese Academy of Sciences), A20 Datun Road, Chaoyang District, Beijing 100101, Peopleʼs Republic of China}
\affiliation{Max-Planck-Institut f\"ur Radioastronomie, Auf dem H\"ugel 69, D-53121 Bonn, Germany}

\author{Lin Wang}
\affiliation{Shanghai Astronomical Observatory, Chinese Academy of Sciences, 80 Nandan Road, Shanghai 200030, Peopleʼs Republic of China}

\author{Yujie Wang}
\affiliation{Shanghai Astronomical Observatory, Chinese Academy of Sciences, 80 Nandan Road, Shanghai 200030, Peopleʼs Republic of China}
\affiliation{College of Astronomy and Space Sciences, University of Chinese Academy of Sciences, Beijing 100049, Peopleʼs Republic of China}

\author{Zhongli Zhang}
\affiliation{Shanghai Astronomical Observatory, Chinese Academy of Sciences, 80 Nandan Road, Shanghai 200030, Peopleʼs Republic of China}
\affiliation{Key Laboratory of Radio Astronomy and Technology (Chinese Academy of Sciences), A20 Datun Road, Chaoyang District, Beijing 100101, Peopleʼs Republic of China}

\author{Zhonghua Feng}
\affiliation{School of Information Engineering, Southwest University of Science and Technology, Mianyang 621010, Peopleʼs Republic of China}

\author{Baoda Li}
\affiliation{College of Physics, Guizhou University, Guiyang 550025, Peopleʼs Republic of China}

\author{Minghui Li}
\affiliation{State Key Laboratory of Public Big Data, Guizhou University, Guiyang 550025, Peopleʼs Republic of China}

\author{Tong Liu}
\affiliation{National Astronomical Observatories, Chinese Academy of Sciences, 20A Datun Road, Chaoyang District, Beijing, 100101, Peopleʼs Republic of China}

\author{Yaowei Li}
\affiliation{College of Physics, Guizhou University, Guiyang 550025, Peopleʼs Republic of China}

\author{Bo Peng}
\affiliation{School of Information Engineering, Southwest University of Science and Technology, Mianyang 621010, Peopleʼs Republic of China}

\author{Yu Pan}
\affiliation{School of Science, Chongqing University of Posts and Telecommunications Chongqing, 40000, Peopleʼs Republic of China}

\author{Yuxiao Wu}
\affiliation{School of Science, Chongqing University of Posts and Telecommunications Chongqing, 40000, Peopleʼs Republic of China}

\author{Liyun Zhang}
\affiliation{College of Physics, Guizhou University, Guiyang 550025, Peopleʼs Republic of China}

\author{Xingnan Zhang}
\affiliation{State Key Laboratory of Public Big Data, Guizhou University, Guiyang 550025, Peopleʼs Republic of China}

\author{Peng Jiang}
\affiliation{National Astronomical Observatories, Chinese Academy of Sciences, 20A Datun Road, Chaoyang District, Beijing, 100101, Peopleʼs Republic of China}
\affiliation{Guizhou Radio Astronomical Observatory, Guizhou University, Guiyang 550025, Peopleʼs Republic of China}
\affiliation{Key Laboratory of Radio Astronomy and Technology (Chinese Academy of Sciences), A20 Datun Road, Chaoyang District, Beijing 100101, Peopleʼs Republic of China}
\affiliation{College of Astronomy and Space Sciences, University of Chinese Academy of Sciences, Beijing 100049, Peopleʼs Republic of China}

\footnote{$\ast$ panzc@bao.ac.cn}
\footnote{$\dag$ hyzhang@bao.ac.cn}
\footnote{$\ddag$ caoshuo@bnu.edu.cn}
\footnote{$^{\S}$ pfreire@mpifr-bonn.mpg.de}

\begin{abstract}

By January 2025, 60 pulsars were discovered by the Five-hundred-meter Aperture Spherical radio Telescope globular cluster (GC) pulsar survey (GC FANS), with spin periods spanning 1.98\,ms to 3960.72\,ms. 
Of these, 55 are millisecond pulsars (MSPs; $P<30$\,ms), while 35 are binaries with orbital periods spanning 0.12\,days to 466.47\,days. 
This paper describes GC FANS, a deep, thorough search for pulsars in 41 GCs in the FAST sky ($-14^\circ < \delta < 65^\circ$) and describes new discoveries in 14 of them. 
We present updated timing solutions for M92A, NGC~6712A, M71A, and M71E, all of which are
``spider'' pulsars with short orbital periods. We present new timing solutions for M71B, C, and D.
With orbital periods of $\sim$ 466 and 378\,days, M71B and M71C are the widest known GC binaries;
these systems resemble the normal wide MSP - He WD systems in the Galactic disk.
With a spin period of 101\,ms, M71D is in an eccentric ($e\sim$0.63) orbit with an 11-day period and a massive companion; the system has a total mass of $2.63 \pm 0.08 \, \Msun$.
These features and its large characteristic age suggest it is a double neutron star system (DNS) formed via massive binary evolution early in the cluster's history, akin to Galactic disk DNSs--unlike other candidate GC DNSs, which typically form dynamically. A comparative analysis of GC pulsar populations within FAST's sky reveals that most clusters (10 of 14) resemble the Galactic disk MSP population, likely due to lower stellar densities.

\end{abstract}


\section{Introduction}
\label{sec:introduction}

\subsection{Pulsars in GCs}

Globular clusters (GCs) generally have high stellar densities of up to $\sim 10^{6}\rm \, \Msun \; pc^{-3}$ in their cores \citep{Baumgardt2018}. 
These immense stellar densities led to a very high rate of interactions between stars (e.g., tidal capture and the exchange encounters of stellar systems), enabling the creation of binary systems, such as low-mass X-ray binaries (LMXBs), where mass and angular momentum of a low-mass star are being transferred to a neutron star (NS).
Therefore, GCs have an unusually large population of LMXBs, which relative to their stellar mass are roughly three orders of magnitude more numerous than in the Galactic disk \citep{Katz1975,Clark1975}.
With spin periods of a few milliseconds, the so-called radio ``millisecond pulsars'' (MSPs) are thought to be the evolutionary products of LMXBs
(\citealt{Alpar_1982,Phinney1994}, for a review see \citealt{Tauris_2023} and references therein). The large number of LMXBs per unit stellar mass in GCs should result in a similarly abundant MSP population. Hence, among targets of directed surveys, GCs have been the most fruitful place to search for MSPs. As of 2024 October, a total number of 343 pulsars have been reported in 45 GCs\footnote{\url{https://www3.mpifr-bonn.mpg.de/staff/pfreire/GCpsr.html}}, with 319 ($\sim$ 93\,\% of the total) being MSPs ($P<30\,\rm ms$)--nearly half the Galactic MSP census from the ATNF catalog\footnote{\url{https://short.url/psrcat}} \citep{Manchester2005}. This contrasts starkly with the Galactic disk, where MSPs constitute $\sim 16\%$ of pulsars, reflecting GCs' ancient ($\sim 10 \rm \, Gyr$) stellar populations.

One feature of MSPs is that, given their evolution, most are in binary systems. Indeed, binary systems account for $\sim$ 56\% in GCs, where the ratio is $\sim$ 12\% in the galaxy as a whole (as reported in the ATNF catalog).
Furthermore, the possibility of repeated exchange interactions in the densest GCs implies that they can produce exotic systems that can not be easily formed in the Galactic disk \citep{Verbunt2014}.
Some unusual GC pulsars include the fastest spinning pulsar PSR J1748$-$2446ad (spin period 1.396\,ms; \citealp{Hessels2006}), which can be used to constrain the equation of state \citep{Cipolletta2015}; 
and especially MSPs with eccentric orbits and massive companions, e.g., PSR J0514$-$4002A (orbital eccentricity 0.89, a companion mass of $1.08 \pm 0.03 \, \Msun$; \citealt{Dutta_2025})
and PSR J1807$-$2500B (orbital eccentricity 0.75, a companion mass $1.207 \pm 0.002 \, \Msun$; \citealt{Lynch2012}).
Unlike MSPs in the Galactic disk, they are thought to form in secondary exchange encounters, i.e., exchange encounters that occur after the MSP was recycled.
Of particular interest are the ``holy grail'' systems, such as (possibly) the first-ever MSP + black hole binary \citep{Barr2024} or even hypothetically an MSP + MSP binary \citep{Faucher-Giguère2011}.

\begin{table*}[ht]
	\centering
	\caption{Survey source list and observation details. Columns 2, 3, 4, and 5 list the heliocentric distance, cluster mass, core radius of the cluster (at which the brightness drops to one-half the central value), and the logarithm of the central luminosity density of the cluster \citep{Harris1996,Harris2010}. The cluster mass is reported in \citet{Baumgardt2018}. Columns 6 and 7 list the approximate upper limit on the pseudo luminosity at 1.4 GHz of an undiscovered pulsar and its typical minimum flux density. Columns 8 and 9 list the single-binary encounter rate and stellar encounter rate. Columns 10 and 11 list the count of isolated and binary pulsars found in the cluster. Column 12 lists the predicted values of the DM in the direction of the cluster. Columns 13 and 14 list the theoretical longest tracking time for the targets within 40$^{\circ}$ of the zenith and the longest single observation length from FAST in hours.} 
	\label{GCs}
	\begin{scriptsize}
	\setlength{\tabcolsep}{1mm}{
            \resizebox{\textwidth}{!}{
		\begin{tabular}{llllllllllllll}
			\hline
                Catalog ID   & $D_{\rm Sun}$ & Mass &  $r_{\rm c}$    &  $\rho_{0}$   &  $L_{\rm max}$$^a$ & $S_{\rm min}$$^b$ & $\gamma_{\rm M4}$ & $\Gamma_{\rm M4}$ & Isolated$^c$ & Binary$^c$ &  Predicted DM$^d$ &  Time Visible & Longest Time   \\
			Name    &   kpc  &  $10^5\,\Msun$  &  arcmin &  log $L_{\odot} \, \rm pc^{-3}$ & mJy kpc$^2$ & $\mu$Jy &  &  & Pulsars  & Pulsars  & $\dmu$   &   Hrs   &   Hrs     \\
			\hline
			NGC 2419      &	82.6	& 7.83  &	0.32	&	1.62	&	10.64&	1.56	& 0.01 &   0.10    &$\cdots$ 	&$\cdots$&	 75        &	6.0	&	5.0	\\
                NGC 4147      &	19.3	& 0.45  &	0.09	&	3.63	&	0.73 &	1.97	& 1.45 &   0.45    &$\cdots$ 	&$\cdots$&	 20        &	5.7	&	3.0	\\
                M53 (NGC 5024)&	17.9	& 5.02  &	0.35	&	3.07	&	0.48 &	1.51	& 0.21 &   0.84    &\textbf{(1)}  &   4\textbf{(3)} &	 20(25.5)   &	5.6	&	5.1	\\
                NGC 5053      &	17.4	& 0.63  &	2.08	&	0.54	&	0.68 &	2.25	& 0.002&   0.005    &$\cdots$ 	&$\cdots$&	 20        &	5.6	&	2.3	\\
                M3 (NGC 5272) &	10.2	& 4.09  &	0.37	&	3.57	&	0.15 &	1.48	& 0.62 &   1.72        &$\cdots$	&  5\textbf{(2)} &	 20(26.4)   &	6.0	&	5.3	\\
                NGC 5466      &	16.0 	& $\cdots$ &1.43	&	0.84	&	0.39 &	1.52	& 0.004&  0.005    &$\cdots$ 	&$\cdots$&	 21        &	6.0	&	5.0	\\
                NGC 5634      &	25.2	& 2.47  &	0.09	&	3.63	&	1.25 &	1.97	& 1.11 &   0.76    &$\cdots$ 	&$\cdots$&	 28        &	3.3	&	3.0	\\
                M5 (NGC 5904) &	7.5	    & 3.92  &	0.44	&	3.88	&	0.09 &	1.67	& 1.02 &   3.83        &1        &   6\textbf{(2)} &	 29(29.5)   &	4.5	&	4.2	\\
                M107 (NGC 6171)& 6.4	& 0.61  &	0.56	&	3.08	&	0.14 &	3.52	& 0.37 &   0.29    &$\cdots$   &$\cdots$&	 88        &	1.4	&	1.0	\\
                M13 (NGC 6205)&	7.1    	& 4.84  &	0.62	&	3.55	&	0.08 &	1.53	& 0.52 &   2.18        &2       &   7\textbf{(4)} &	 31(30.3)   &	6.0	&	5.0	\\
                M12 (NGC 6218)  &	4.8 & 1.06  &	0.79	&	3.23    &	0.04 &	1.81	& 0.42 &   0.54       &$\cdots$ 	&   \textbf{(2)} &53(42.6)    &	4.0	&	3.6 \\
                NGC 6229      &	30.5	& 2.47  &	0.12	&	3.54	&	1.55 &	1.67	& 0.62 &   1.46    &$\cdots$ 	&$\cdots$&	 31        &	5.7	&	4.2	\\
                M10 (NGC 6254)&	4.4	    & 1.89  &	0.77	&	3.54	&	0.04 &	1.98	& 0.67 &   1.25        &\textbf{(2)}	    &$\cdots$&	 80(43.7)   &	3.7	&	3.0	\\
                M92 (NGC 6341)&	8.3	    & 2.73  &	0.26	&	4.30    &	0.11 &	1.53	& 2.53 &   6.99       &$\cdots$ 	&   \textbf{(2)} &	 35(35.4)   &	5.9	&	5.0	\\
                NGC 6366      &	3.5	    & 0.47  &	2.17	&	2.39	&	0.03 &	2.59	& 0.08 &   0.12    &$\cdots$ 	&$\cdots$&	 128       &	3.5	&	2.0	\\
                M14 (NGC 6402)&	9.3	    & 5.95  &	0.79	&	3.36	&	0.17 &	2.02	& 0.25 &   3.15        &$\cdots$ 	&  \textbf{(5)} &	 126(80.5)  &	3.8	&	3.0	\\
                NGC 6426      &	20.6	& 0.73  &	0.26	&	2.47	&	0.75 &	1.76	& 0.12 &   0.08    &$\cdots$ 	&$\cdots$&	 88        &	4.6	&	4.0	\\
                NGC 6517      &	10.6	& 2.16  &	0.06	&	5.29	&	0.30 &	2.70	& 26.82 &  18.54        &20\textbf{(17)}	    &   1 &	 214(182.1) &	2.7	&	2.5	\\
                NGC 6535      &	6.8	    & 0.20  &	0.36	&	2.34	&	0.16 &	3.56	& 0.23 &   0.01    &$\cdots$ 	&$\cdots$&	 101       &	4.2	&	1.0	\\
                NGC 6539      &	7.8	    & 2.15  &	0.38	&	4.15	&	0.16 &	2.56	& 1.55 &   7.85    &$\cdots$ 	&$\cdots$&	 183       &	3.0	&	2.8	\\
                NGC 6712      &	6.9	    & 0.95  &	0.76	&	3.18	&	0.13 &	2.72	& 0.29 &   0.86        &$\cdots$ 	&   \textbf{(1)} &	 185(155.1) &	2.8	&	2.0	\\
                NGC 6749      &	7.9	    & 2.05  &	0.62	&	3.30    &	0.69 &	11.07	& 0.35 &   1.14    &$\cdots$ 	&$\cdots$&	 289       &	4.5	&	3.0	\\
                NGC 6760      &	7.4	    & 2.86  &	0.34	&	3.89	&	0.12 &	2.13	& 1.35 &   2.30    &$\cdots$ 	&$\cdots$&	 182       &	4.4	&	4.0	\\
                M56 (NGC 6779)&	9.4	    & 1.66  &	0.44	&	3.28	&	0.14 &	1.59	& 0.41 &   0.76    &$\cdots$ 	&$\cdots$&	 101       &	6.0	&	5.0	\\
                M71 (NGC 6838)&	4	    & 0.38  &	0.63	&	2.83	&	0.03 &	1.62	& 0.40 &   0.06        &$\cdots$ 	&   5\textbf{(4)} &	 102(116.9) &	5.7	&	5.0	\\
                NGC 6934      &	15.6	& 1.50  &	0.22	&	3.44	&	0.47 &	1.91	& 0.59 &   0.91    &$\cdots$ 	&$\cdots$&	 67        &	5.0	&	3.3	\\
                M72 (NGC 6981)&	17	    & 0.81  &	0.46	&	2.38	&	0.99 &	3.43	& 0.08 &   0.12    &$\cdots$ 	&$\cdots$&	 40        &	1.7	&	1.0	\\
                NGC 7006      &	41.2	& 1.32  &	0.17	&	2.58	&	3.05 &	1.80	& 0.11 &   0.19    &$\cdots$ 	&$\cdots$&	 60        &	5.7	&	3.7	\\
                M15 (NGC 7078)&	10.4	& 5.18  &	0.14	&	5.05	&	0.22 &	2.00	& 8.89 &   42.42        &   14\textbf{(7)}	&   1 &	 44(66.8)   &	5.3	&	3.0	\\
                M2 (NGC 7089) &	11.5	& 6.24  &	0.32	&	4.00	&	0.31 &	2.31    & 1.05 &   7.21        &$\cdots$ 	&  \textbf{(10)} &	 35(43.6)   &	4.1	&	2.2	\\
                GLIMPSE-C01   &	4.2   	& $\cdots$ &0.59	&	$\cdots$&	0.16 &	8.97	& $\cdots$&  $\cdots$&$\cdots$ 	&   6\textbf{(1)} &	 257(486.7) &	4.0	&	1.0	\\
                IC 1257       &	25	    & 1.80  &	0.25	&	$\cdots$&	1.67 &	2.68	& $\cdots$&  $\cdots$&$\cdots$ 	&$\cdots$&	 196       &	3.1	&	3.0	\\
                Pal7 (IC 1276)&	5.4  	& 0.74  &	1.01	&	2.78	&	0.07 &	2.28	& 0.17 &  0.23    &$\cdots$ 	&$\cdots$&	 164       &	3.1	&	3.0	\\
                Ko 1          &	48.3	& $\cdots$&	0.33	&	$\cdots$&$\cdots$ &	$\cdots$& $\cdots$&  $\cdots$&$\cdots$ 	&$\cdots$&	 21        &	5.3	&	$\cdots$	\\
                Ko 2          &	34.7	& $\cdots$&	0.25	&	$\cdots$&$\cdots$ &	$\cdots$& $\cdots$&  $\cdots$&$\cdots$ 	&$\cdots$&	 77        &	5.9	&	$\cdots$	\\
                Pal 2         &	27.2	& 2.20  &	0.17	&	4.06	&	1.40 &	1.90	& 0.90 &   14.00    &$\cdots$ 	&$\cdots$&	 181       &	6.0	&	5.0	\\
                Pal 3         &	92.5	& 0.19  &	0.41	&	0.01	&$\cdots$& 	$\cdots$& 0.001 &  0.0008     &$\cdots$	&$\cdots$&	 32        &	4.2	&	$\cdots$	\\
                Pal 4         &	108.7	& 0.15  &	0.33	&	0.11	&$\cdots$& 	$\cdots$& 0.001 &  0.001     &$\cdots$	&$\cdots$&	 18        &	6.0	&	$\cdots$	\\
                Pal 5         &	23.2	& 0.13  &	2.29	&	-0.58	&	1.06 &	1.97 	& 0.0004&  0.0002      &$\cdots$ 	&$\cdots$&	 30        &	4.2	&	3.0	\\
                Pal 10        &	5.9	    & 1.25  &	0.81	&	3.51	&	0.13 &	3.83	& 0.46 &   2.24    &$\cdots$ 	&$\cdots$&	 230       &	5.7	&	2.7	\\
                Pal 11        &	13.4	& 0.14  &	1.19	&	2.29	&	0.49 &	2.70	& 0.03 &   0.37    &$\cdots$	&$\cdots$&	 90        &	3.0	&	1.7	\\
                Pal 13        &	26	    & 2.78  &	0.48	&	0.16	&	1.03 &	1.53	& 0.004 &  0.0001     &$\cdots$ 	&$\cdots$&	 30        &	5.4	&	5.0	\\
                Pal 14        &	76.5	& 0.19  &	0.82	&	-0.97	&	22.35&	3.82	& 0.0002 & 0.00007      &$\cdots$ 	&$\cdots$&	 31        &	5.5	&	0.8	\\
                Pal 15        &	45.1	& 0.53  &	1.2   	&	--0.25	&	7.87 &	3.87	& 0.0005 & 0.0007      &$\cdots$ 	&$\cdots$&	 63        &	4.2	&	0.8	\\
                Whiting 1     &	30.1	& 1.37  &	0.25	&	$\cdots$&	2.82 &	3.11	& $\cdots$ &  $\cdots$   &$\cdots$ 	&$\cdots$&	 25        &	3.8	&	1.2	\\
			\hline
	\end{tabular}}}
    \end{scriptsize}
	\begin{itemize}
            \item[a] We assume $L_{1400}=S_{1400}D_{\rm Sun}^{2}$, where $S_{1400}$ is taken from $S_{\rm min}$. The pulsar is assumed to be isolated and has a spin period $\ge$ 1 ms and a $S/N$ of 10.
            \item[b] $S_{\rm min}$ is estimated from Eq.~\ref{eq:fluxestimate}, where $T_{\rm obs}$ is taken from the longest single observation time for each cluster in GC FANS.
            \item[c] The values in parentheses denote the number of pulsars that were found by FAST (including M5G, M13F, and M71E that were not first discovered in GC FANS).
            \item[d] Predicted DM represents the predicted values of the DM that are based on the YMW16 electron-density model of the Galaxy (\citet{Yao2017}), while values in parentheses indicate the average DM derived from the known pulsars in the GCs. There is no formal uncertainty on these values. Since these GC pulsars have been applied to refine the YMW16 model itself, some predicted DM values are consistent with the DM of pulsars in the cluster (e.g., M5). The predicted DM values differ from the true values by a factor between 0.66 to 1.83, which shows the characteristic discrepancy between the values.
            \label{table:targets}
            
	\end{itemize}
\end{table*}

\subsection{Surveys for pulsars in GCs}

The unusual characteristics of the GC MSP population have motivated many surveys.
The first successful survey was carried out with the Lovell 76-m radio telescope (5 pulsars; e.g., \citealp{Lyne1988}); later with the Murriyang (Parkes) 64-m radio telescope ($\sim$ 48 pulsars; e.g., \citealp{Camilo2000}), the Arecibo 305-m radio telescope ($\sim$ 28 pulsars; e.g., \citealp{Hessels2007}) and the Green Bank Telescope (GBT) ($\sim$ 83 pulsars; e.g., \citealp{Ransom2005,Freire2008,Lynch2011,Lynch2012,DeCesar2015}). 

One fundamental limitation of all these surveys has always been the sensitivity of the radio telescopes. Other considerations, like a field of view or survey speed, generally do not apply because the regions of interest of GCs represent a very small area in the sky, which in many cases (especially single dish telescopes) can be covered with a single, or a few, beams. Sensitivity is the main limitation because radio pulsars are intrinsically faint radio sources, typically found at distances of the order of $\sim \, 1$ kpc. The GCs associated with our Galaxy have distances one order of magnitude larger, which means that, for most GCs, we only detect the very brightest pulsars: the vast majority remain undetected \citep{Turk2013}. This implies that, whenever more sensitive radio telescopes become available, more pulsars in GCs become detectable. This is currently the case, where three new, more sensitive instruments have, in the last 6 years, doubled the number of pulsars in GCs.
The upgraded giant metrewave radio telescope (uGMRT) has found 9 GC pulsars\footnote{Seven of them are from the GCGPS survey, all in GCs without previously known pulsars, see \url{http://www.ncra.tifr.res.in/~jroy/GC.html}} (e.g., \citealp{Freire2004,Gautam2022,Gupta2017}) and the South African 64-dish MeerKAT radio telescope array has found 105 pulsars (e.g., \citealp{Ridolfi2021,Ridolfi2022,Chen2023}\footnote{These surveys, which benefit greatly from MeerKAT's Southern location, are conducted under the TRAPUM project, see \url{http://trapum.org/discoveries/}}). In what follows, we describe the searches made with the Five-hundred-meter Aperture Spherical radio Telescope (FAST; \citealp{Nan2011,Jiang2019,Jiang2020}). 

\subsection{FAST survey for pulsars in GCs}

With an effective illuminated aperture of 300\,m in diameter and a 19-beam L-band receiver, FAST is the most sensitive single-dish radio telescope in the world.
Within its declination ($\delta$) range of $-14\degr$ to $+65\degr$, FAST achieves 2-3 times higher sensitivity and double the sky-coverage of Arecibo, making it ideal for deep pulsar surveys. 
To date, FAST has discovered over 1000 Galactic pulsars, constituting $\sim$ 25\% of the known Galactic pulsar population (e.g., CRAFTS, \citealp{Li2018}; GPPS, \citealp{Han2021,Han2024,Wang2024}; initial GC pulsar survey, \citealp{Pan2021b}). The FAST GC pulsar survey, \textbf{G}lobular \textbf{C}luster with \textbf{F}AST: \textbf{A} \textbf{N}eutron-star \textbf{S}urvey (GC FANS), leverages FAST's high sensitivity to conduct a thorough L-band searching of GC pulsars within FAST's sky, regardless of their stellar interaction rate--a good predictor of the number of pulsars in a GC. A pilot survey, Search of Pulsars in Special Populations (SP$^2$), utilized an ultra-wide-band (UWB) receiver covering 270-1620\,MHz to observe the GC M92 in October 2017 \citep{Pan2020}. Since May 2018, monitoring of GCs within FAST's sky has employed the 19-beam L-band receiver covering 1050-1450\,MHz \citep{Pan2021b}.
GC FANS aims to search for pulsars that may be missed by previous surveys due to sensitivity limitations and to characterize extreme and exotic systems.
Such discoveries will advance our understanding of GC dynamics, constrain exotic stellar evolution pathways, and probe neutron star physics \citep{Hessels2015}.

In this work, we present an overview of GC FANS. In Section~\ref{sec:observations}, we provide details of over seven years of FAST observations and outline the data reduction process.
In Section~\ref{sec:results}, we summarize discoveries in 14 clusters.
Particularly, we provide a detailed study and updated timing solutions for PSR~J1717+4308A in M92 (henceforth M92A), PSR~J1853$-$0842A in NGC 6712 (henceforth NGC~6712A), and PSRs J1953+1846A and E in M71 (henceforth, M71A and E), along with new timing solutions for PSRs J1953+1846B, C, and D (henceforth, M71B, C, and D). M71D will merit special discussion, as it is likely a double neutron star system (DNS) coeval with its host GC. In Section~\ref{sec:discussion}, we make a general statistical analysis of the GC pulsar sample established through FAST. Finally, we give a brief summary of our findings and an outlook on future studies in Section~\ref{sec:conclusions}.

\section{Survey Details}
\label{sec:observations}

\begin{figure*}
\centering
	\includegraphics[width=17cm,height=7.5cm]{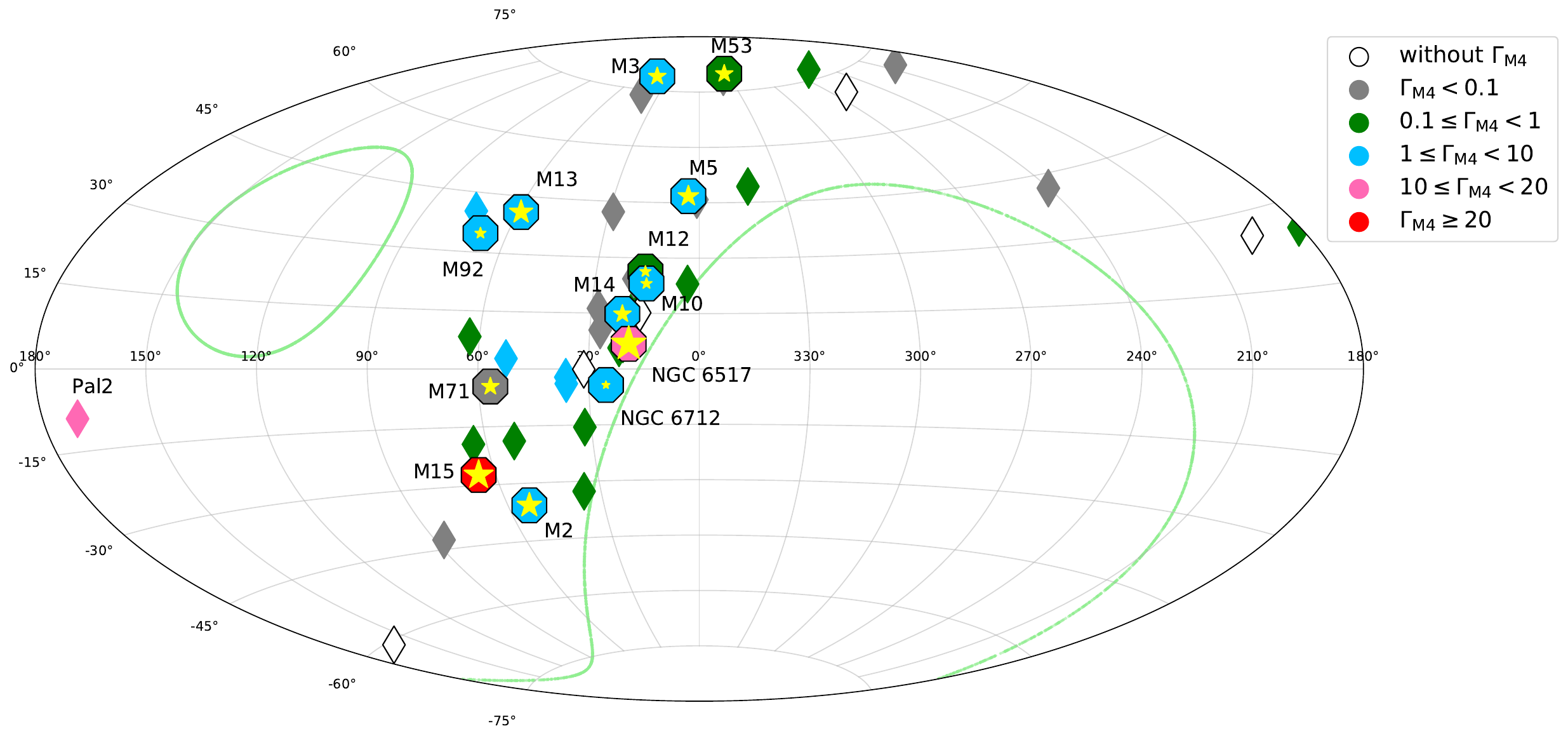}
	\caption{The distribution of all the 45 GCs with its $\Gamma_{\rm M4}$ (listed in Table~\ref{table:targets}) within the FAST sky (delineated by green lines) in Galactic coordinates. Hexagons and diamonds are denoted as GCs with and without known pulsars, respectively. The symbols in different colors represent different values of $\Gamma_{\rm M4}$, while the five GCs without a measurement of $\Gamma_{\rm M4}$ are denoted as white diamonds. The higher $\Gamma_{\rm M4}$ generally correlates with a higher possibility of finding pulsars. For the 14 clusters having pulsar discoveries, the size of the yellow star at each of these points corresponds to the number of known pulsars in that cluster.}
        \label{fig:pulsarnum}
\end{figure*}

\begin{figure*}
\centering
	\includegraphics[width=16cm,height=12cm]{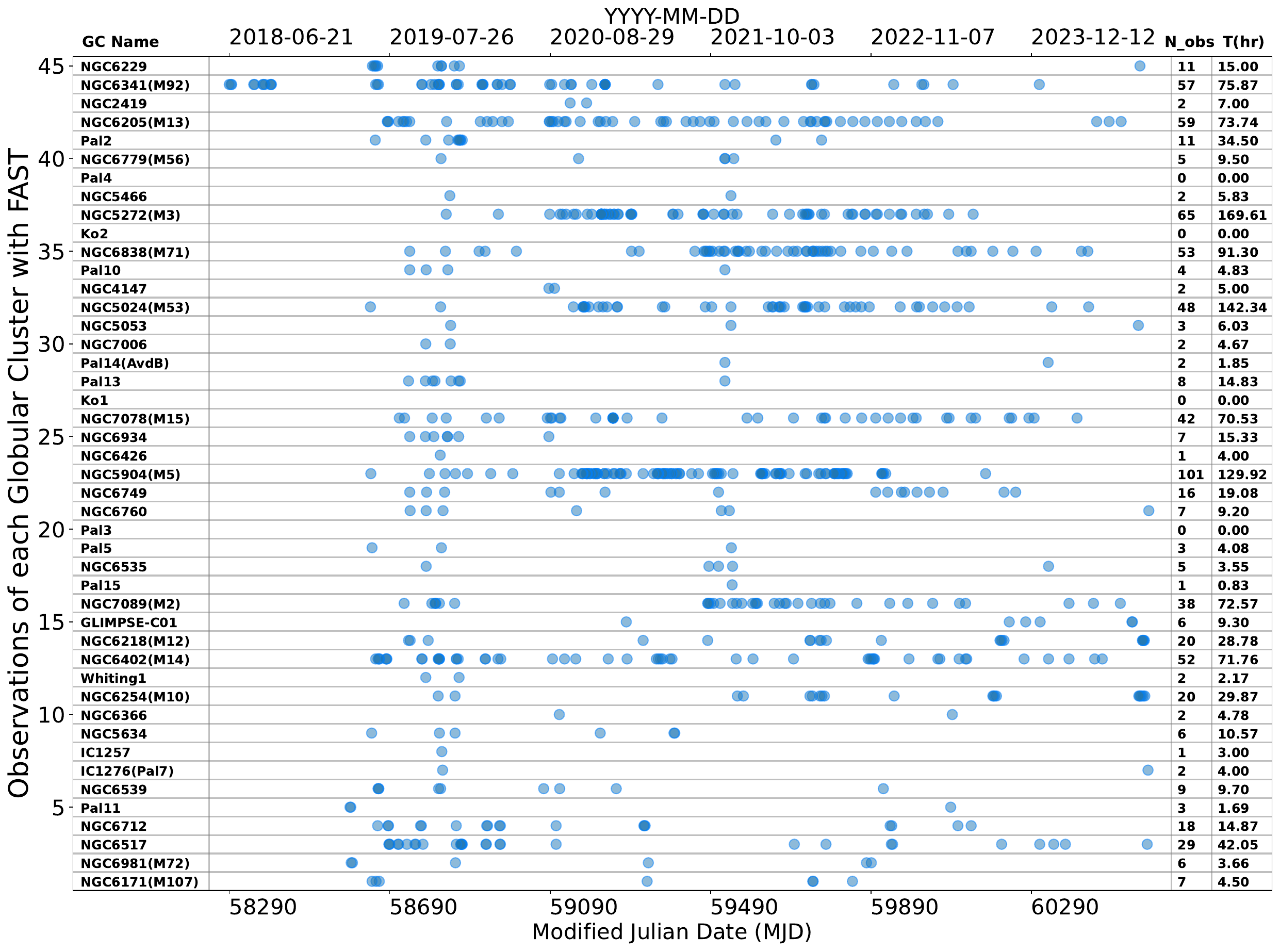}
	\caption{Summary of GC FANS observations for 45 clusters within $\rm -13.1^{\circ}< DEC < +47.5^{\circ}$ from May 2018 to September 2024, plotted as a function of Modified Julian Date (MJD) and calendar date (YYYY-MM-DD). Clusters are sorted by their declinations (north to south). Blue markers denote individual observations, with the observation counts (N\_obs) and total integration time (T) in hours shown on the right. Clusters with pulsar discoveries (e.g., M3) are regularly monitored, while those without pulsar detections are observed less frequently to optimize observing time.}
        \label{fig:observationdata}
\end{figure*}

\begin{figure*}
\centering
	\includegraphics[width=13cm,height=8cm]{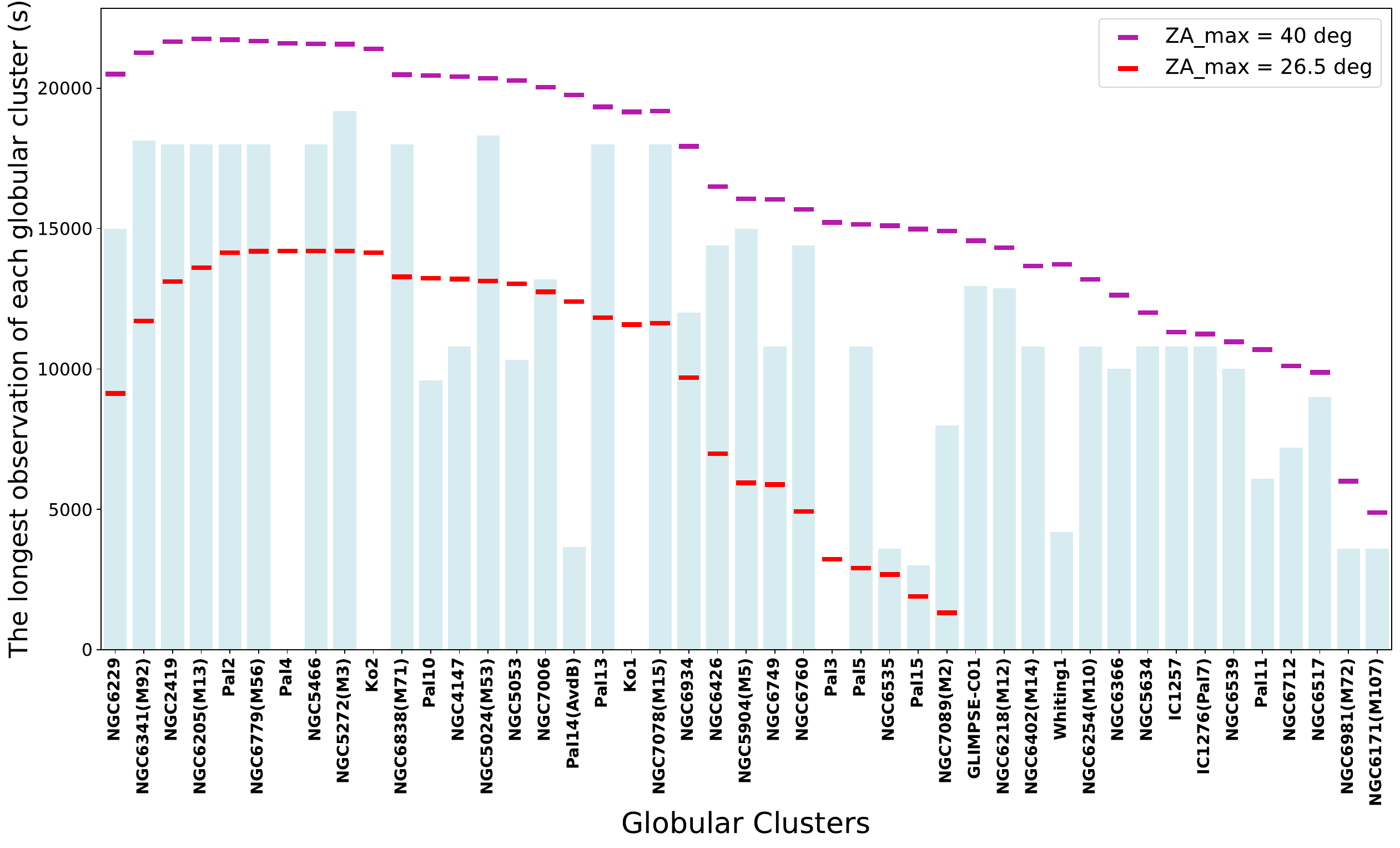}
	\caption{The summary of the longest single observation durations for all 45 clusters within FAST's sky. Red and purple horizontal lines denote the theoretical maximum observation times at zenith angles of 26.5$^{\circ}$ and 40$^{\circ}$, respectively. Clusters are ordered by their declinations (north to south). Within a zenith angle of 26.5$^{\circ}$, FAST maintains a stable gain ($\sim 16\,\rm K\,Jy^{-1}$) with a 300-meter effective aperture, but beyond this angle, backlighting effects reduce the effective aperture to as low as 200 meters, resulting in decreased gain.}
        \label{fig:longesttime}
\end{figure*}

\begin{figure}
	\includegraphics[width=8.6cm,height=7cm]{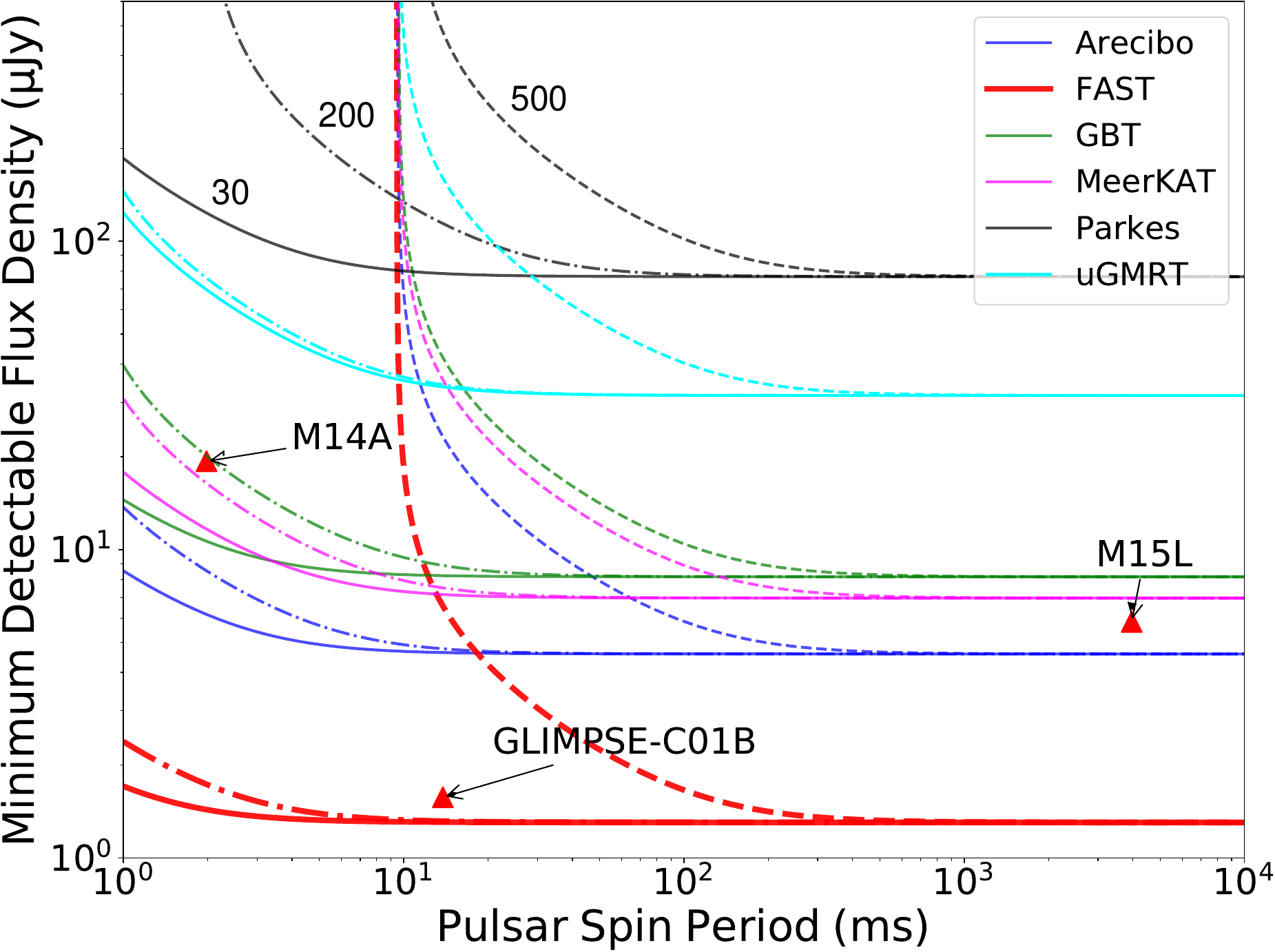}
	\caption{Survey sensitivity as a function of period and DM, assuming an intrinsic pulse width of 8\%. Each set of curves shows (from left to right, see example denoted for the curves of Parkes) the sensitivity calculated using DMs of 30, 200, and 500\,$\dmu$. In each case, the DM = 30\,$\dmu$ curve is the leftmost and the DM = 500\,$\dmu$ curve is the rightmost. To illustrate real detection performance, we plotted the mean flux density for three extreme cases--GLIMPSE-C01B, M14A, and M15L--calculated via Equations~\ref{eq:fluxestimate}-\ref{eq:scatt} using their measured DM and spin period from Table~\ref{table:pulsars} and the integration times of 4 hr (S/N$\sim$3), 3 minutes (S/N$\sim$15), and 5 hr (S/N$\sim$50), respectively.}
    \label{fig:sensitivity}
\end{figure}

\subsection{Targets}
\label{sec:targets}

Of the 45 Galactic GCs visible to FAST ($\rm -13.1^{\circ}< DEC < +47.5^{\circ}$),
41 within 83\,kpc of the Sun were selected to observe by GC FANS.
{The remaining 4 clusters (Ko 1, Ko 2, Pal 3, and Pal 4) are low-priority targets due to their large distances (e.g., $D_{\rm sun} = 108.7 \, \rm kpc$ for Pal 4).
The fundamental and derived parameters for all 45 GCs are presented in Table~\ref{table:targets}, including counts of isolated and binary pulsars. FAST-discovered pulsars are highlighted in bold (parenthetical values).

Following \citet{Verbunt2014}, we calculate the stellar encounter rate $\Gamma_{\rm M4} \, \propto \, \rho_{\rm c}^2 r_{\rm c}^3/v_0$ and the encounter rate per binary, $\gamma_{\rm M4} \, \propto \, \rho_{\rm c} /v_0$, normalized to the GC M4.
Here, $\rho_{\rm c}$, $r_{\rm c}$, and $v_0$ denote core density, core radius, and central velocity dispersion, respectively.
Both $\gamma_{\rm M4}$ and $\Gamma_{\rm M4}$ are listed in Table~\ref{table:targets}.
In Fig.~\ref{fig:pulsarnum}, we map the Galactic distribution of these GCs, color-classified by $\Gamma_{\rm M4}$, which correlates with the number of X-ray binaries in GCs and serves as an indicator of potential pulsar population (i.e., a higher $\Gamma_{\rm M4}$ suggests a larger number of pulsars; \citealt{Bahramian2013}).
It should be noted that all the 41 targeted GCs within 83\,kpc were selected without any bias against low-$\Gamma$ clusters.
In Section~\ref{sec:discussion}, we further discuss the correlation between the stellar interaction and pulsar discoveries in GCs within the FAST sky.

\subsection{Observing Strategy and Survey Limitations}
\label{sec:Observing Strategy}

The Modified Julian Date (MJD) timeline for FAST observations of all 45 clusters (including the four clusters that have not been observed yet) is shown in Fig.~\ref{fig:observationdata}.
The complete observational records for each cluster are available online\footnote{\url{https://github.com/lyjastro/GC-FANS.git}}, and the latest observation catalog can be found in FAST's website\footnote{\url{https://fast.bao.ac.cn/}}.

The GC FANS pilot survey began in October 2017 with an observation for M92, using a UWB receiver covering 270-1620\,MHz with a $\sim$60\,K system temperature \citep{Pan2020}.
From May 2018 onward, we started monitoring observations at FAST, using the 19-beam L-band receiver, covering a frequency range of 1.05-1.45\,GHz.
Data were recorded using 8-bit sampling for two or four polarizations, with 4096 channels (0.122 MHz channel width), and a sampling time of 49.152 $\mu \rm s$. 
All observations were stored into standard \texttt{PSRFITS} files in pulsar search mode. We note that coherent de-dispersion was not available in the observing system during our campaign, and thus all pulsar data were processed using incoherent de-dispersion. Since pulsars are typically weak radio sourves and GCs are often distant from the sun ($D_{\rm Sun} > 5\,\kpc$ for $\sim$91\% of known GCs), clusters with known pulsars (e.g., NGC 6517) and, in most cases, with a distance to the Sun $D_{\rm Sun} < 10 \, \kpc$ have been chosen to be observed first. In Fig.~\ref{fig:longesttime}, we show the actual and theoretical longest single FAST observation times for each cluster. To maximize the search sensitivity, most targets were observed for the full predicted integration time at a zenith angle of 26.5$^{\circ}$. Only four clusters (Pal 10, NGC 4147, NGC 5053, Pal 14, and Ko 1) received shorter exposures, corresponding to 73\%, 82\%, 77\%, and 24\% of their theoretical maxima, respectively.

Pulsars in GCs are found to centrally concentrate, typically lying within $r/r_{\rm c}\lesssim 3$ (where $r_{\rm c}$ is the cluster core radius); a result of mass segregation of massive NSs \citep{Davies1997}. The central beam size of FAST at 1.4\,GHz is roughly 3$'$, which should effectively cover the central regions of 32 out of 45 GCs ($\rm r_{c} \lesssim 0.5'$) within FAST's sky, enabling detection of nearly all pulsars with a single pointing. Six clusters (Pal 5, NGC 6366, NGC 5053, NGC 5466, Pal 15, and Pal 11) have extended core radii ($r_{\rm c} > 1.2'$), comparable to or exceeding the FAST beam half-power radius ($\sim 1.5'$), which could result in 20–50\% of pulsars falling outside the beam. Despite this limitation, we prioritized the central beam observations using the 19$-$beam receiver with TRACKING and SWIFTCALIBRATION modes for computational efficiency, pointing at the central coordinates of the 41 clusters \citep{Harris2010}.

\subsection{Survey Sensitivity}
\label{sec:sensitivity}

The minimum detectable flux density sensitivity, $S_{\rm min}$, can be estimated \citep[see, e.g.,][]{Lorimer2004} as a function of the radiometer noise for a pulsar with period $P$ and observed pulse width $W_{\rm obs}$ as 
\begin{equation}
S_{\rm min} = \beta \displaystyle  \frac{(S/N_{\rm min}) T_{\rm sys}}{G \sqrt{n_{\rm p} T_{\rm obs} \Delta f}}\sqrt{\frac{W_{\rm obs}}{P-W_{\rm obs}}},
\label{eq:fluxestimate}
\end{equation}
where $\beta \sim 1$ is the sampling efficiency for our 8-bit recording system, the minimum signal-to-noise ratio ($S/N)_{\rm min}$ is 10, the system temperature ($T_{\rm sys}$) is 24~K, the antenna gain ($G$) is 16\,K\,Jy$^{-1}$, and the number of polarizations ($n_{\rm p}$) is 2 \citep{Pan2021b}. 
$T_{\rm obs}$ is the integration time and varies between our target clusters from 0.8-5.1\,hr depending on the declination of the source.
$\Delta f$ is the bandwidth in MHz, taken to be 400 MHz here from 1.05\,GHz to 1.45\,GHz. 

The observed pulse width, $W_{\rm obs}$, can be expressed \citep[see, e.g.,][]{Hessels2007} in 
terms of the intrinsic pulse width $W_{\rm int}$, the finite time sampling of the data recorder $t_{\rm samp}$, the dispersion smearing across individual frequency channels $t_{\rm DM}$, the dispersion smearing due to the finite dispersion measure (DM) step size of the time series $t_{\rm \Delta DM}$, and the interstellar medium (ISM) in the form of scattering $t_{\rm scatt}$ as
\begin{equation}
W_{\rm obs} = \sqrt{W_{\rm int}^2 + t_{\rm samp}^2 + t_{\rm DM}^2 + t_{\rm \Delta DM}^2 + t_{\rm scatt}^2}.
\end{equation}
The dispersive smearing across an individual channel 
\begin{equation}
t_{\rm DM} = 8.3\left(\frac{{\rm DM}}{{\rm
pc~cm^{-3}}}\right)\left(\frac{\Delta\nu_{\rm chan}}{{\rm MHz}}\right)\bigg(\frac{\nu_{\rm center}}{{\rm GHz}}\bigg)^{-3}\mu{\rm s}.
\label{eq:3}
\end{equation}
The smearing due to an incorrect DM in the time series 
\begin{equation}
t_{\Delta{\rm DM}} = 4.1\left[\left(\frac{\nu_{\rm low}}{{\rm GHz}}\right)^{-2} -
\left(\frac{\nu_{\rm high}}{{\rm GHz}}\right)^{-2}\right] \left(\frac{\Delta{\rm
DM}}{{\rm pc~cm^{-3}}}\right){\rm ms},
\end{equation}
where $\nu_{\rm low}$ and $\nu_{\rm high}$ are the low and high-frequency edges of the bandwidth, respectively. Following \cite{Cordes2022}, the scattering time
\begin{eqnarray}
\label{scatt.eqn}
\tau_{\rm scatt} &=& 1.9 \times 10^{-7}~{\rm ms}~ \left(\frac{{\rm DM}}{{\rm pc}~{\rm cm}^{-3}}\right)^{1.5} \nonumber \\
&& \times \left[ 1 + 3.55 \times 10^{-5} \left( \frac{{\rm DM}}{{\rm pc}~{\rm cm}^{-3}}\right)^{3.0} \right].
\label{eq:scatt}
\end{eqnarray}

The basic observational system parameters of GC pulsar surveys for each telescope are summarized in Table~\ref{table: Telescopes basic parameters}, used to estimate $W_{\rm obs}$.
For $\Delta{\rm DM}$, the general way is applying the DDplan.py routine, from the PulsaR Exploration and Search TOolkit \citep[{\sc presto}; ][]{Ransom2001,Ransom2002,Ransom2003}, to calculate the optimum DM step size and down-sampling factor, then evaluating the DM deviation.
This step contributes to the dispersion smearing due to the incorrect DM trial across the whole band.
Here we adopt a maximum DM spread ($\Delta{\rm DM}$) between the pulsar's DM and that assumed in the de-dispersed trial time series as 0.05\,$\dmu$, to quantity $t_{\Delta{\rm DM}}$ and perform the comparison of survey sensitivities.
Using these parameters and assuming a 4-hour integration time for all six telescopes, we present the sensitivity curves in Fig.~\ref{fig:sensitivity} estimated from Equations~\ref{eq:fluxestimate}-\ref{eq:scatt}, illustrating how sensitivity depends on DM and pulsar spin period.
For $\rm DM \leq 200 \, \dmu$, FAST is sensitive to slow and mildly recycled pulsars ($P = 30 \sim 100 \, \rm ms$) down to a minimum flux density of $S_{\rm min} \sim 1.30 \, \rm \mu Jy$ at 1.4\,GHz; for MSPs with $P \leq 10$\,ms, the broadening interstellar scattering raises $S_{\rm min}$, particularly at $\rm DM > 200 \, \dmu$.
In this figure, we also compare GC FANS to major GC pulsar surveys (Arecibo, MeerKAT, GBT, Parkes, and uGMRT) at DMs of 30, 200, and 500\,$\dmu$, with uGMRT's Band 4 receiver ($\nu_{\rm center} \sim 650$~MHz) and other L-band receivers ($\nu_{\rm center} \sim 1.4$~GHz).

To illustrate both the theoretical sensitivity limits and flux densities of actual detections--even when using different integration times for extreme pulsars--we displayed the measured mean flux densities (using Eq.~\ref{eq:fluxestimate}) of the pulsar with the highest DM PSR J1848$-$0129B (hereafter GLIMPSE-C01B; $\rm DM \sim \rm 482 \, \dmu$; 4-hour integration, S/N$\sim$3; Li et al. in preparation), the fastest spinning pulsar PSR J1737$-$0314A (hereafter M14A; $P \sim 1.98 \; \rm ms$; 3-minute integration, S/N$\sim$15; \citealt{Pan2021b}), and the slowest pulsar PSR J2129+1210L (hereafter M15L; $P \sim 3961 \; \rm ms$; 5-hour integration, S/N$\sim$50; \citealt{Wu2024}), among discoveries in GC FANS.

\begin{table*}[!htpb]
    \centering
    \caption{Basic observational system parameters of GC Pulsar Surveys. Column 2 lists the sampling efficiency that accounts for various sensitivity losses due to signal processing and digitization. Columns 3 and 4 list the system temperature and antenna gain. Columns 5 and 6 list the bandwidth and central observing frequency. Column 7 lists the number of channels. Column 8 lists the low- and high-frequency edges of the bandwidth.}
    \setlength{\tabcolsep}{4mm}{
        \begin{tabular}{cccccccc} 
            \hline
            Telescope & $\beta$ & $T_{\rm sys}$ & $G$ & $\Delta f$ & $\nu_{\rm center}$ & Channels & Frequency range  \\
                      &         &    K &   $\rm K\,Jy^{-1}$ & MHz  &  GHz &  &  GHz \\
            \hline
            FAST$^1$      &  1   & 24  &  16    &  400 &  1.25 & 4096  & 1.05-1.45         \\
            Arecibo$^2$   &  1.2 & 40  &  10.5  &  300 &  1.4 & 256 (512)$^a$  & 1.15-1.55  \\
            MeerKAT$^3$   &  1.1 & 26  &  2.8   &  650 &  1.284 & 1024  & 0.96-1.60         \\
            GBT$^4$       &  1.05& 22.8 &  2    &  650 &  1.5 &  512 & 1.18-1.83           \\
            uGMRT$^5$     &  1   & 102.5 & 4.2  &  180 &  0.65 &  4096 & 0.56-0.74           \\
            Parkes$^6$    &  1.5 & 35  &  0.7   &  288 &  1.374 & 96  & 1.23-1.52          \\
            \hline
        \end{tabular}}
        
    \begin{itemize}
        \item[]
    \textbf{Notes:} 1. FAST \citep{Pan2021b}; 2. Arecibo \citep{Hessels2007}; 3. MeerKAT \citep{Ridolfi2021}; 4. GBT \citep{Martsen2022}; 5. uGMRT \citep{Gautam2022,Das2025}; 6. Parkes \citep{Camilo2000}; a: Following \cite{Hessels2007}, we consider 256 channels for DM$<$100\,$\dmu$ and 512 channels for DM$>$100\,$\dmu$ to ensure sufficient resolution.
    \end{itemize}
    \label{table: Telescopes basic parameters}
\end{table*}

\subsection{Data Reduction}
\label{sec:data reduction}

\subsubsection{Pulsar Searching}

The searching analysis was conducted using {\sc presto}. Initial radio frequency interference (RFI) mitigation was performed with \texttt{rfifind}, which automatically flagged contaminated frequency channels and time segments. Additionally, some strongly polluted channels were manually removed; these included channels 0-200 (frequency range 1.04-1.07\,GHz), 640–810 (1.13-1.15\,GHz), and 3800-4095 (1.51-1.55\,GHz) in certain observations. For GCs with known pulsars, data were initially dedispersed around the pulsar's average DM using a $\pm10\%$ DM window with either \texttt{prepdata} or \texttt{prepsubband}. The DM range was then extended based on the empirical relation between the average DM and DM spread \citep{Yin2023} for clusters hosting pulsars with DM $>100\,\dmu$ (e.g., GLIMPSE-C01).
For clusters without detected pulsars, dedispersion was performed with \texttt{prepsubband} over a DM range spanning 2-3 times the predicted value calculated from the YMW16 model, ensuring a thorough search across DM space.

We carried out periodicity searches in both the Fourier and time domains across all observations conducted by GC FANS: the Fourier domain acceleration search, based on the Fast Fourier Transform (FFT; \citealt{Ransom2003}), and the time domain search, based on the Fast Folding Algorithm (FFA; \citealt{Staelin1969}).
In the Fourier domain, dedispersed time series were Fourier-transformed with \texttt{realfft} and searched using \texttt{accelsearch} (summing up to 32 harmonics).
The parameter $z_{\rm max}$ in \texttt{accelsearch} defines the maximum number of Fourier bins a signal can drift through due to factors like orbital motion or pulsar spin-down during the observation (see \citealt{Ransom2002}).
We set $z_{\rm max}=20$ (covering --20 to +20 Fourier-bin drifts, including $z = 0$) to allow for mild accelerations from the cluster potential or hidden long-period companions when searching for isolated pulsars, thereby maximizing detection yield with minimal added computation cost; for binary searches we set $z_{\rm max}$ up to 1200 to accommodate the orbital motions \citep{Pan2021b}.
However, It has been noticed that the above assumption is not valid when $\Delta t_{\rm obs} \gtrsim 0.1P_{\rm b}$, where $\Delta t_{\rm obs}$ is the observation length and $P_{\rm b}$ is the binary orbital period \citep{Ransom2003}.
Consequently, we initially searched the entire data for each observation to maximize the signal-to-noise ratio ($S/N$). Considering the computational cost, we then segmented the data (using 10–30\% of the original integration time) to search for possible compact binaries in GCs with detected pulsars using a $z_{\rm max}=1200$ (e.g., \citealt{Pan2021b}).
Furthermore, We also applied the ``jerk search'' functionality of \texttt{accelsearch} \citep{Andersen2018} for highly accelerated binaries, primarily in M5 and M53, which host previously known binaries \citep{Pan2021b}.
In the time domain, we used the FFA code Riptide$-$FFA\footnote{\url{https://riptide-ffa.readthedocs.io/en/latest/index.html}} \citep{Morello2020}, over a search period ranging from $4 \, \rm ms$-$100\,\rm s$ for most GCs (see details in \citealt{Li2025}), and then extended to 0.1\,s$-$180\,s for core-collapsed clusters within FAST's sky (NGC 6517 and M15) to search for slow pulsars (see e.g., \citealt{Wu2024}).
Besides, we also employed the stack search (e.g., \citealt{Pan2016,Cadelano2018}) for NGC 6517 and M15, which involves incoherently combining power spectra from all available observations and amplify faint signals buried below the noise, to search for weak periodic signals (see details in Dai et al. in preparation). 

Candidate sifting was mainly performed by the \texttt{ACCEL\_sift.py} routine from {\sc presto} and \texttt{Jinglepulsar} \footnote{\url{https://github.com/jinglepulsar}} to combine multiple detections of each pulsar candidate from \texttt{accelsearch} results.
We also identified any candidates recurring on different days with similar spin periods and DM values (Yu et al. in preparation).
For the FFA and stack search, candidates with a peak above 7$\sigma$ and 5$\sigma$, respectively ($\sigma$ being the standard deviation of the folded time series and the spectral powers), were saved into a candidate file.
All the sifted candidates were first folded with \texttt{prepfold} on the dedispersed time series to eliminate obvious false signals; every candidate that passed this initial check was then folded on the raw \texttt{PSRFITS} data. The resulting diagnostic plots were inspected visually.

\subsubsection{Timing}

In this work, we used the \texttt{fitorb.py} routine\footnote{\url{https://github.com/scottransom/presto/blob/master/bin/fitorb.py}} to fit the initial orbital parameters for the new binary pulsars (M71B to D). For each detection, we derived the pulse time of arrivals (ToAs) with \texttt{get\_TOAs.py} by crossing-correlating the standard pulse profile against a high-$S/N$ template, which was generated from fitting a set of Gaussian curves (using \texttt{pygaussfit.py} from {\sc presto}) to the best detection.
For the subsequent analysis of ToAs, we used \texttt{TEMPO}\footnote{\url{http://tempo.sourceforge.net}} to derive the phase-connected timing solutions \citep{Freire2018}.
With the latter program, we inserted the initial estimated orbital parameters into the pulsar ephemeris and utilized \texttt{sieve} script for new pulsars (M71B to D) to look for a single phase offset (so-called ``JUMP'') between epochs and \texttt{DRACULA}\footnote{\url{https://github.com/pfreire163/Dracula.}} script for known pulsars (M92A, NGC 6712A, and M71A and E), which could automatically determine JUMPs between epochs.
As for the previously known eclipsing spider pulsars (M92A, NGC 6712A, and M71A), we excluded TOAs obtained during eclipse phases (see the caption of Fig.~\ref{fig:timingresiduals}) in the timing analysis, avoiding the excess delays that can bias the updated timing solutions.

\begin{table*}[ht]
	\centering
	\caption{Basic parameters of GC pulsars discovered by FAST include dispersion measure (DM), spin period ($P$), spin-period derivative ($\dot{P}$), orbital period ($P_{\rm b}$), eccentricity ($e$), and the companion mass ($M_{\rm c}$), assuming a 1.35\,$\Msun$ pulsar and an inclination angle of $i=60^\circ$. Pulsars marked with “i” are isolated, while others are in binaries, including 15 systems with orbital parameters still under analysis. GC FANS identified 25 isolated and 35 binary pulsars. Additionally, \emph{M5G}, \emph{M13F}, and \emph{M71E} are also included in this table, which were not first found in GC FANS.}
	\label{GC pulsars}
	\begin{scriptsize}
	\setlength{\tabcolsep}{0.8mm}{
		\begin{tabular}{lllllll|lllllll}
                \hline
			Name &  DM     & $P$ & $\dot{P}$   &  $P_{\rm b}$ &  e & $M_{\rm c}$ & Name &  DM     & $P$ & $\dot{P}$   &  $P_{\rm b}$ &  e & $M_{\rm c}$ \\
			              & ($\dmu$)&   (ms)         & $10^{-20}\rm s ~ s^{-1}$ &  (days)        &             &    $\Msun$ &   & ($\dmu$)&   (ms)         & $10^{-20}\rm s ~ s^{-1}$ &  (days)        &             &    $\Msun$   \\
			\hline
			M53B$\rm ^{a,b}$&	25.96	       &	6.242	&	--1.746	&47.67735	&	0.01		&	0.35 	 	    	   &  NGC 6517O$\rm ^{h}$&	182.49	&	4.287	&     12.139	&	i	    &	i	&	i	\\
                M53C$\rm ^{a,b}$&	26.19	       &12.535	        &	5.26		&	i	    	&	i	        &	i		           	   & NGC 6517P$\rm ^{h}$&	183.05	&	5.536	&$\cdots$	&	i	    &	i	&	i	\\
                M53D$\rm ^{a,b}$&	24.61	       &	6.069	&	2.26		&	5.75024	&	0.000013	&	0.30            	   & NGC 6517Q$\rm ^{h}$&	182.45	&	7.258	&$\cdots$	&	i &	i	&	i	\\
                M53E$\rm ^{a,b}$&	25.88	       &	3.972	&	--0.758	&	2.43138	&	0.000009	&	0.21 	  	   & NGC 6517R$\rm ^{h}$&	182.50	&	6.585	&$\cdots$	&	i	    &	i	&	i	\\
                M3E$\rm ^{a,c}$&	26.52	       &	5.473	        &	--0.105	&	7.09685	&	0.000309&	0.23 		    & NGC 6517S$^{\ast}$&	182.50     &	3.77	&	$\cdots$ &   i	  &	i		&	i	 \\
                M3F$\rm ^{a,c}$&	26.44	       &	4.404	        &	3.1224		&	2.99199	&	0	        &	0.17 		    & NGC 6517T$^{\ast}$&	182.50     &	3.68	&	$\cdots$ &   i	  &	i		&	i       \\
                M5F$\rm ^{a,d}$&	29.41	       &	2.654		& 2.21		&	1.60952	&	0	        &	0.23 	    & NGC 6517U$^{\ast}$&	 183.90	    &	6.02	&	$\cdots$   &   i	  &	i	&	i     \\
                \emph{M5G}$\rm ^{d}$&	29.40	   &	2.75		&1.247		&0.11393	&	0	&	0.022   			            & NGC 6517V$^{\ast}$ &	 182.90	    &	4.55	&	$\cdots$   &   i	  &	i		&	i           \\
                \emph{M13F}$\rm ^{e}$&	30.37     &	3.004	&	1.40	        &	1.378	&	0		&	0.16	 	    	            &  J1848-0129B$^{\ast}$&	 481.95    &	13.801	&	$\cdots$   &   $\cdots$	  &	$\cdots$		&	$\cdots$    \\
                M13G$^{\ast}$&	 30.75    &	4.323	&	$\cdots$&	$\cdots$	&	$\cdots$&	$\cdots$    			    & NGC 6712A$\rm ^{i}$&	  155.13    &	2.149	&	--0.240      &0.14828	&	0		&	0.03     \\   
                M13H$^{\ast}$&	 30.75  &	11.21	&	$\cdots$	&	$\cdots$	&	$\cdots$		&	$\cdots$    & M71B$\rm ^{a,j}$&	119   	&	 79.899	&100.7	&	466.46888	&	0.0019693	& 0.49 \\
                M13I$^{\ast}$&	 29.45   &	6.37	&	$\cdots$	 &	$\cdots$	&	$\cdots$		&	$\cdots$             & M71C$\rm ^{a,j}$&	116.20	&	28.933	&4.04	&	378.23276	&	0.0003775	& 0.41	  \\
                M12A$^{\ast}$&	42.50	 &	2.36		&$\cdots$		&$\cdots$		&	$\cdots$	&	$\cdots$  	    & M71D$\rm ^{a,j}$&	118.60	&	100.679	&8.0	&	10.93804	&	0.628921	& 1.63	 \\
                M12B$^{\ast}$&	42.70	&	2.76		&$\cdots$		&$\cdots$		&	$\cdots$	&	$\cdots$     & \emph{M71E}$\rm ^{k,l}$&	113.10	&	4.444	&	2.217	&	0.03704	&	0.0005	&	0.008	\\
                M10A$\rm ^{a}$&	43.90    &	4.73		&$\cdots$		&$\cdots$		&	$\cdots$	&	$\cdots$     &  M15I$\rm ^{a}$&	67.50	&	5.122	&$\cdots$	&   i	  &	i		&	i	\\
                M10B$\rm ^{a}$&	43.36	       &	7.35		&$\cdots$		&$\cdots$		&	$\cdots$	&	$\cdots$   &  M15J$\rm ^{m}$&	66.69	&	11.843	& $\cdots$	&  i	&	i	&	i	\\
                M92A$\rm ^{f}$&	35.45	       &	3.160	&	6.1185	&	0.20087	&	$\cdots$	&	0.18 	             & M15K$\rm ^{m}$&	66.50	&	1928.451	&$\cdots$	&i	&	i	&	i	\\
                M92B$^{\ast}$&	 35.35    &		3.51  &	$\cdots$		&	2.29        &	0	&	0.24			      & M15L$\rm ^{m}$&	66.10	&	3960.716	&$\cdots$	&i	&	i	&	i	 \\
                M14A$\rm ^{a}$&	82.10   &	1.980	&	9.53		&	0.2278	&	0		&	0.018 				      & M15M$^{\ast}$&		67.86	&	 	4.836       &$\cdots$	&i	&	i	&	i	 \\
                M14B$\rm ^{a}$&	81  	&	8.52		&$\cdots$		&$\cdots$		&	$\cdots$	&	$\cdots$   		& M15N$^{\ast}$&		66.60	&		9.289	&$\cdots$	&i	&	i	&	i	\\
                M14C$\rm ^{a}$&	80	     &	8.46		&$\cdots$		&$\cdots$		&	$\cdots$	&	$\cdots$   	& M15O$\rm ^n$&		67.44	&		11.066	&--219.46	&i	&	i	&	i	\\
                M14D$\rm ^{a}$&	78.80	&	2.89		&$\cdots$		&	0.7428	&	0		&	0.142 				& M2A$\rm ^{a}$&	43.30	&	10.15	&	--30.8	&	4.2555	&	0.0751	&	0.19	\\
                M14E$\rm ^{a}$&	80.40	&	2.28		&$\cdots$		&	0.8463	&	0		&	0.195				& M2B$\rm ^{a}$&	43.80	&	6.97	&	--26 	&	9.3471	&	0	&	0.21	\\
                NGC 6517E$\rm ^{g}$&	183.18     &	7.602	&	--103.598	&	i   		&	i		&	i 					& M2C$\rm ^{a}$&	44.10	&	3.00	    &	--3.15	&	1.1091	&	0	&	0.16	\\
                NGC 6517F$\rm ^{g}$&	183.79   &24.892		&	--267.08	&	i   		&	i		&	i 					& M2D$\rm ^{a}$&	43.60	&	4.22	&	7.81	&	3.4297	&	0	&	0.27	\\
                NGC 6517G$\rm ^{g}$&	185.06      &51.591		&	10.7	&	i   		&	i		&	i 				& M2E$\rm ^{a}$&	43.80	&	3.70	    &$\cdots$	&$\cdots$	&	$\cdots$	&	$\cdots$\\
                NGC 6517H$\rm ^{a,h}$&	179.63      &	5.643	&	3.431	&	i	    	&	i		&	i 					& M2F$^{\ast}$&	43.40	&	4.78	&$\cdots$	&$\cdots$	&	$\cdots$&	$\cdots$\\
                NGC 6517I$\rm ^{a,h}$&	177.88	&	3.254	&	--1.642	&	i	    &	i	&	i				  				& M2G$^{\ast}$&	43.40	&	2.535	&$\cdots$	& 0.12036	& 0	& 0.015 \\
                NGC 6517K$\rm ^{h}$&	182.38	&	9.591	&   81.80	&	i	    &	i	&	i				  				& M2H$^{\ast}$&	43.20	&	2.878	&$\cdots$	&$\cdots$	&	$\cdots$&	$\cdots$\\
                NGC 6517L$\rm ^{h}$&	185.74	&	6.057	&    --3.38	&	i	    &	i	&	i				  				& M2I$^{\ast}$&	44.10	&	8.78	&$\cdots$	&$\cdots$	&	$\cdots$&	$\cdots$\\
                NGC 6517M$\rm ^{h}$&	183.18	&	5.357	&   48.586 	&	i	    &	i	&	i		          					& M2J$^{\ast}$&	43.5	0&	4.50	&$\cdots$	&$\cdots$	&	$\cdots$&	$\cdots$\\
                NGC 6517N$\rm ^{h}$&	182.64	&	4.995	&   --53.15	&	i	    &	i	&	i	&   &   &  &  &  &  &   \\
			\hline
	\end{tabular}}
    \end{scriptsize}
	\begin{itemize}
            \item[]\textbf{References}: a \citet{Pan2021b}; b \citet{Lian2023}; c \citet{Li2024}; d  \citet{Zhang2023a}; e \citet{Wang2020}; f \citet{Pan2020}; g \citet{Pan2021a}; h \citet{Yin2024}; i \citet{Yan2021}; j this paper; k \citet{Han2021}; l \citet{Pan2023}; m \citet{Wu2024}; n \citet{Dai2025}; $^\ast$ Independent publications presenting details on these 18 pulsars are currently in preparation.
            \end{itemize}
    \label{table:pulsars}
\end{table*}

\begin{figure}
	\includegraphics[width=\columnwidth]{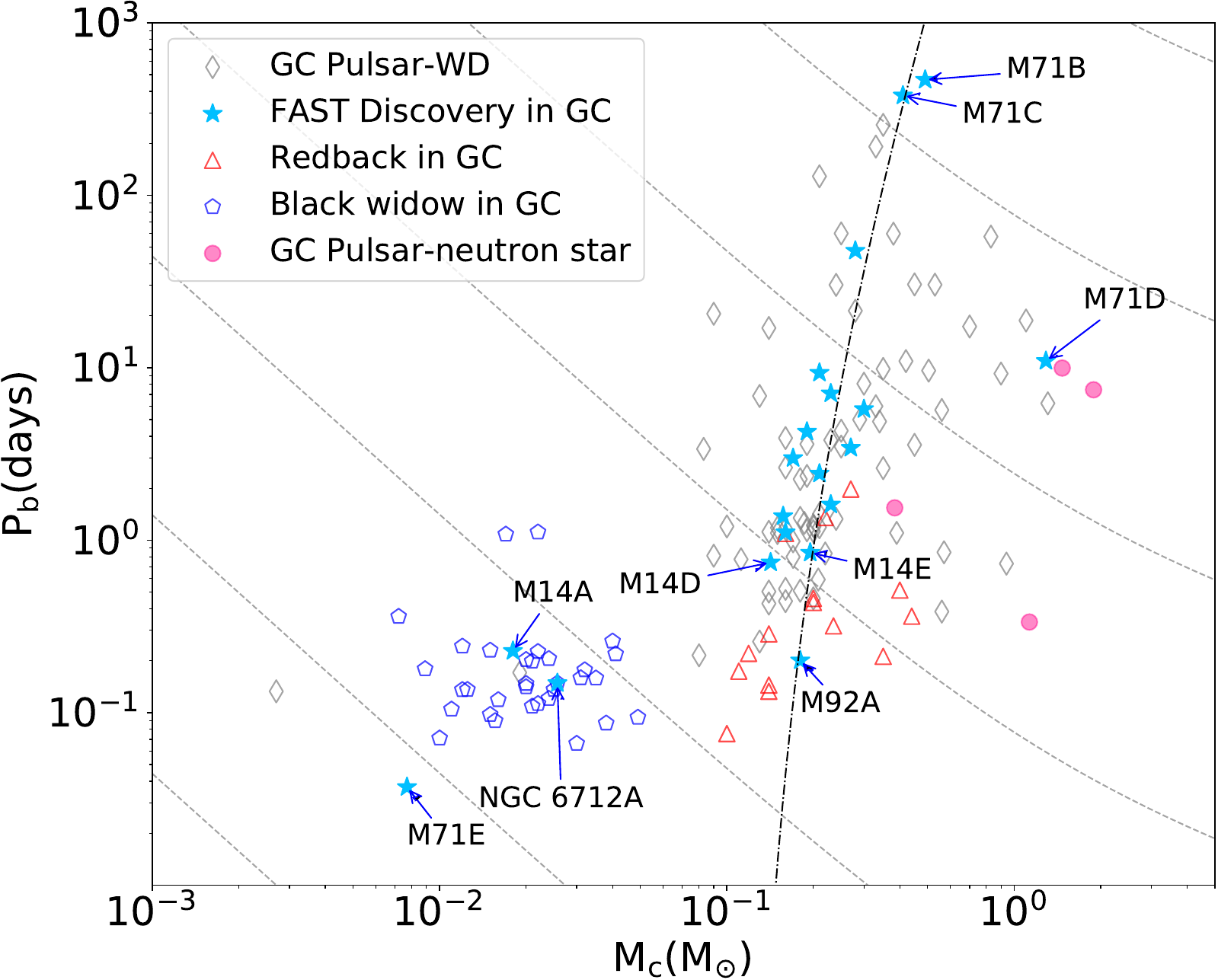}
	\caption{Companion mass vs. orbital period for GC pulsars. Pulsars in GCs from the ANTF pulsar catalog are shown as unfilled grey diamonds, known black widows and redbacks are unfilled pentagons and triangles respectively, and FAST discoveries are filled stars. The dashed grey lines show a constant semi-major axis $a = x / \sin i$ (here we assumed a pulsar mass of 1.35\,$\Msun$ in the mass function). The line in the lower left corner denotes $a = 1$~light ms and each successive line denotes an increase by a factor of 10. The dash-dotted black line shows the numerical estimation of the non-conservative evolution of close binary systems with low-mass ($1-2 \; \Msun$) donor stars that were transferring material to the primary \citep{Tauris1999}. }
    \label{fig:Pb_mc}
\end{figure}

\begin{table*}
\begin{center}
\caption{Observed and derived parameters of M92A and NGC 6712A. Eccentricity and the longitude of periastron for these two pulsars are set quantities (labeled by $s$). The orbital model used is the BTX model \citep{Shaifullah2016}.}
\label{table:timing6712}
\begin{tabular}{lll}
\hline\hline
Pulsar         &   J1717+4308A    &   J1853$-$0842A      \\
\hline
Start of Timing Data (MJD)                       &              58351.440         &   58659.715                 \\
End of Timing Data (MJD)                            &          60021.946      &   60617.410                     \\
Reference Epoch (MJD)                            &          58390.000000    &   58685.702931                         \\
Number of TOAs            &  1438    &   475          \\
EFAC &      2.18    &        1.38                     \\
Residuals RMS ($\mu$s)     &   18.49     &   3.79     \\
Reduced $\chi^2$   &     1.010     &     1.007                        \\
Solar System Ephemeris       &   DE440                         &                DE440                          \\
Binary Model          &   BTX            &              BTX                                           \\
\hline
\multicolumn{2}{c}{Measured quantities}  \\
\hline
Right Ascension, $\alpha$ (J2000)               &   17:17:06.4964(1)  & 18:53:04.07403(2)   \\
Declination, $\delta$ (J2000)                &   43:08:03.4786(9)      & -08:42:28.264(1)    \\
Proper Motion in $\alpha$, $\mu_{\alpha}$ (mas\,yr$^{-1}$)    &  $-$4.3(8)  & 1.1(2)  \\
Proper Motion in $\delta$, $\mu_{\delta}$ (mas\,yr$^{-1}$)    & $-$1.3(4)  & 5.5(4)  \\
Spin Frequency, $f_{\rm 0}$ (s$^{-1}$)          &        316.48368670265(5)      &   465.23897161367(2)         \\
First Spin Frequency Derivative, $f_{\rm 1}$ (s$^{-2}$)   &  $-$6.1284(7)$\times 10^{-15}$   &   5.200(7)$\times 10^{-16}$ \\
Second Spin Frequency Derivative, $f_{\rm 2}$ (s$^{-3}$)  &  ---  &   $-$4.3(8)$\times 10^{-26}$  \\
Dispersion Measure, DM (pc cm$^{-3}$)                 &         35.4205(4)        &   155.1407(2)                \\
Projected Semi-major Axis, $x_{\rm p}$ (lt-s)   &        0.398705(1)     &   0.0491828(3)            \\
Orbital Eccentricity, $e$    &   0.00$^s$                       &           0.00$^s$                     \\
Longitude of Periastron, $\omega$ (deg) &   0.00$^s$             &          0.00$^s$             \\
Time of passage at Periastron, $T_0$ (MJD)     &     58353.5490817(1) &   58685.6416204(2)          \\
Orbital Frequency, $f_{\rm b}$ (s$^{-1}$)       &      5.762033597(5)$\times 10^{-5}$     &   7.80539542(1)$\times 10^{-5}$        \\
First Orbital Frequency Derivative, $f_{\rm b,1}$ (s$^{-2}$)   &  $-$1.31(2)$\times 10^{-19}$  &   $-$8.7(5)$\times 10^{-20}$ \\
Second Orbital Frequency Derivative, $f_{\rm b,2}$ (s$^{-3}$)    &  1.67(3)$\times 10^{-27}$ &   $-$1.01(6)$\times 10^{-27}$ \\
\hline
\multicolumn{2}{c}{Derived Parameters}  \\
\hline
Spin Period, $P$ (s)            &        0.0031597205227817(3)      &   0.0021494330032833(1)              \\
First Spin Period Derivative, $\dot{P}$ (s s$^{-1}$)     &   6.1185(7)$\times 10^{-20}$   &   $-$2.402(3)$\times 10^{-21}$ \\
Mass Function, $f(M_{\rm p}, M_{\rm c})$ ($\rm M_{\odot}$)&  $1.69\times10^{-3}$  & $5.81\times10^{-6}$  \\
Minimum Companion Mass, $M_{\rm c, min}$ ($\Msun$)  & 0.1601 &  0.0227  \\
Angular Offset from Center in $\alpha$, $\theta_{\alpha}$ (arcmin) & 0.1630  &  $-$0.0558 \\
Angular Offset from Center in $\delta$, $\theta_{\delta}$ (arcmin) & $-$0.0987 &  $-$0.1044\\
Total Angular Offset from Ccenter, $\theta_{\perp}$ (arcmin)        & 0.1906  &  0.1184 \\ 
Total Angular Offset from Center, $\theta_{\perp}$ (core radii)    &  0.7329 &  0.1558\\ 
Projected Distance from Center, $r_{\perp}$ (pc)                   &  0.4601 &  0.2376\\
\hline
\end{tabular}
\end{center}
\end{table*}

\begin{table*}

 \caption{Timing solutions for previously discovered pulsars M71A to E.
            The binary models used are the ELL1 model \citep{Lange2001}, from which the eccentricity $e$ is derived for M71A, B, C, and E (labeled by $a$), and the DD model \citep{Damour1985}.
        \label{table:timing}}   
	     \begin{footnotesize}
         \rotatebox{90}{
         \begin{tabular}{lccccc}
				\hline\hline
				Pulsar name  & M71A   &      M71B       &      M71C      &      M71D  &   M71E    \\
				\hline
				MJD range   &    58741---60302   & 58741---60302   & 58741---60302  &  58741---60302 &  58741---60302  \\
    			Reference epoch (MJD)  &   52812.0000  &   59579.2426  &  59579.2426  & 59579.2426 & 59474.6305 \\
				Number of ToAs         &     1685      &     197       &    178       &  162       &   810     \\
                    EFAC       &     1.64        &     1.010    &     1.086   &  1.005    &  1.109      \\
				Timing residual r.m.s. ($\mu$s) &  10.64  & 141.33 & 123.16 & 204.02 & 37.23 \\
                    Reduced $\chi^2$        &     1.010           &     1.010    &     1.004   &  1.010     &   1.006     \\
				Solar System ephemeris model  &  DE440  &  DE440  &  DE440   & DE440   &  DE440   \\
				Binary model    &   ELL1   &   ELL1   &   ELL1   &  DD  &  ELL1   \\    
				\hline
				\multicolumn{4}{c}{Measured quantities} \\
				\hline
				Right Ascension, $\alpha$ (J2000) &  19:53:46.41603(1)   &  19:53:50.348(1)     &  19:53:51.642(7)   &  19:53:46.8298(8) & 19:53:37.94666(8) \\
				Declination, $\delta$ (J2000)     &  +18:47:04.7796(3)   &  +18:48:07.60(3)     &  +18:48:05.5(4)    &  +18:47:47.04(2) & +18:44:54.317(1) \\
                Spin Frequency, $f_{\rm 0}$ (s$^{-1}$)    & 204.57006473144(5)   &  12.515752038663(9)  &  34.56285133892(3)    & 9.93255637407(1)  & 225.01840471097(2) \\
		      First Spin Frequency Derivative, $f_{\rm 1}$ (s$^{-2}$) &  $-2.03401(8) \times 10^{-15}$   &  $-1.578(5) \times 10^{-16}$    &  $-4.82(6) \times 10^{-17}$  & $-7.9(3) \times 10^{-18}$  & $-1.1224(7) \times 10^{-15}$ \\
    			Dispersion measure, DM (pc~cm$^{-3}$)   & 117.334(2) & 119.17(6) & 117.55(6) & 118.4(1) & 113.10(2) \\
				Orbital Period, $P_{\rm b}$ (days)      &  0.17679503185(7)  &  466.46888(2)   &  378.23276(6)  & 10.9380426(1)  & 0.0370398633(3) \\
				Projected Semi-major Axis, $x$ (lt-s)  & 0.0782158(9)      &  165.83245(3)   &  126.1360(6)  & 32.0634(1)     & 0.006667(2) \\
				Orbital Eccentricity, $e$              & 0.000110(8)$^{a}$           &   0.0019693(4)$^a$    &  0.0003775(4)$^a$    & 0.628921(3)     & 0.0005(6)$^a$   \\
				Longitude of Periastron, $\omega$ (deg) & --- &   ---     &  ---  & 60.4651(4)  & ---    \\
                Rate of advance of Periastron, $\dot{\omega}$ (deg $\rm yr^{-1}$) & --- &   ---   &  ---   & 0.0116(2) &  ---    \\
 				Time of passage at Periastron, $T_0$ (MJD) &  --- &  --- &  ---  & 59525.083519(8)   & --- \\
				Time of passage at Ascending Node, $T_\textrm{asc}$ (MJD)   &    52811.876196(2)  &  59576.58684(2)  &    59510.06378(5)  & ---  & 58829.260068(6)  \\
				First Laplace-Lagrange Parameter, $\epsilon_1 = e \sin\omega$  & $1.0(2) \times 10^{-4}$ &  $1.5167(4) \times 10^{-3}$  & $2.696(4) \times 10^{-4}$  &  --- & $5(6) \times 10^{-4}$ \\
				Second Laplace-Lagrange Parameter, $\epsilon_2 = e \cos\omega$  & $-5.0(11) \times 10^{-5}$ &  $-1.256(3) \times 10^{-3}$   & $-2.640(4) \times 10^{-4}$  & ---  & $0.00(5) \times 10^{-6}$ \\
				\hline
				\multicolumn{4}{c}{Derived quantities} \\
				\hline
    		Spin Period, $P$ (s)  & 0.004888300745824(1)  &   0.07989931383355(6)  &    0.02893279811304(3)   &    0.1006790157880(1) & 0.0044440809243335(4)  \\
				First Spin Period Derivative, $\dot{P}$ (s s$^{-1}$) &  $4.8604(2)\times 10^{-20}$ & $1.007(3)\times 10^{-18}$  &  $4.04(4)\times 10^{-20}$ &  $8.0(3)\times 10^{-20}$ & $2.217(1)\times 10^{-20}$  \\
                    Mass Function, $f(M_{\rm p}, M_{\rm c})$ ($\Msun$)  & 0.0000164364(6)  &   0.02250551(2)    &   0.0150613(2)  &   0.296092(4) &  0.0000002320(2) \\  
				Minimum Companion Mass, $M_{\rm c, min}$ ($\Msun$) &  0.0323   &   0.4210   &   0.3600   &   1.2888  &  0.0077 \\
                    Angular Offset from Center in $\alpha$, $\theta_{\alpha}$ (arcmin)  &  $-$0.0175 &   $-$0.9134 &  $-$1.2189 &   $-$0.0807 &  +2.0224  \\
                    Angular Offset from Center in $\delta$, $\theta_{\delta}$ (arcmin)  &  +0.3279   &   +1.3755 &  +1.3432 &   +1.0325 &  $-$1.8465 \\
                    Total Angular Offset from Center, $\theta_{\perp}$ (arcmin)         &  0.3284    &   1.6512  &  1.8139 &   1.0356  & 2.7385 \\ 
                    Total Angular Offset from Center, $\theta_{\perp}$ (core radii)     &  0.5213    &   2.6209  &  2.8791 &   1.6438  & 4.3468 \\ 
                    Projected Distance from Center, $r_{\perp}$ (pc)                    &  0.3821    &   1.9212  &  2.1105 &   1.2050  & 3.1864 \\
				\hline
		\end{tabular}}
	\end{footnotesize}
\end{table*}

\section{Results}
\label{sec:results}

\subsection{Summary of Pulsars Discovered in GC FANS}

To date, GC FANS has discovered 60 pulsars in 14 clusters (see details in Table~\ref{table:pulsars}), including the first discoveries in M2, M10, M12, M14, M92, and NGC 6712. About 67\% of the total discoveries reside in clusters located beyond 10\,kpc from the Sun, where the sensitivity of FAST gives us a real advantage. 
Notably, we found 17 pulsars in NGC 6517 (e.g., \citealt{Yin2024}), which now hosts the largest pulsar population in the FAST sky, 21 pulsars. Among other pulsars discovered in GCs with previously known pulsars, 4 are located in M53, the most distant GC with known pulsars ($\sim$17.9\,kpc; \citealt{Pan2021b,Lian2023}). Nineteen pulsars, in a total of seven clusters, were recently confirmed and will be described in detail in forthcoming studies. The majority of these discoveries (50 pulsars) were first identified through Fourier-domain acceleration searches (e.g., \citealt{Pan2021b}), while four pulsars (M15J, M15K, M15L as reported in \citealt{Wu2024} and M13I in \citealt{Li2025}) were detected via the FFA, and six pulsars (NGC 6517S to V, M15M, and M15N) were discovered using the stack search method (see Dai et al. in preparation).

The pulsars discovered by GC FANS exhibit spin periods spanning 1.98\,ms (M14A) to 3.96\,s (M15L, the longest-period GC pulsar).
Fifty$-$five of them ($\sim92\%$ of the discoveries) are MSPs with periods $P \lesssim 30$\,ms, including 27 pulsars with $P \lesssim 5$\,ms. 
Thirty$-$four are binary pulsars, with orbital periods ranging from 0.12\,days (M2G) to 466\,days (M71B). In Fig~\ref{fig:Pb_mc}, we display the orbital period–companion mass ($P_{\rm b} - M_{\rm c}$) distribution of all GC binaries, highlighting two ``black widow" (BWs; M14A and NGC 6712A), three ``redback" (M14D, M14E, and M92A) pulsars found in GC FANS, which are compact and eclipsing binaries (for a review, see \citealt{Roberts2013} and references therein). Notably, we found a possible DNS system (M71D) that is likely coeval with its host cluster and two widest GC binaries (M71B and M71C). 

GC FANS also confirmed two previously discovered pulsars: M3A, first discovered in the 1.4\,GHz Arecibo GC pulsar survey \citep{Hessels2007} and later confirmed by \citet{Li2024}, and M71E, discovered by the FAST GPPS survey \citep{Han2021} and confirmed by \citet{Pan2023}. Both are BW pulsars; particularly, M71E has the shortest orbital period among pulsar binaries ($P_{\rm b}=53\,\rm minutes$).
Also, combining the FAST GC FANS observations of M53 with earlier Arecibo observations, we solved a 35$-$year timing solution for PSR B1310+18A (M53A), one of the widest binaries in GC ($P_{\rm b}=256\,\rm days$); this in turn allowed the identification of its companion in Hubble Space Telescope data (see \citealt{Lian2025}). In what follows, we discuss in more detail pulsar discoveries where we have either updated published timing solutions, or obtained new ones.

\subsection{Properties of the Pulsars in M92}

M92A is a 3.15-ms pulsar in a ``redback" system with an orbital period of 4.8 hr. It is located 0.73 core radii from the center (see Fig.~\ref{fig:xray}). Its X-ray counterpart has been identified and exhibits a luminosity of $8.3 \times 10^{31}\,\rm erg\,s^{-1}$ in the 0.3–8\,keV band \citep{Zhao2022}. Following the observations until May 2020 reported by \cite{Pan2021b}, we carried out 14 additional FAST observations from August 2020 to March 2023. The updated timing solution for M92A is presented in Table~\ref{table:timing6712}, with the corresponding postfit timing residuals shown in Fig.~\ref{fig:timingresiduals}. The extended 3-yr baseline allows us to measure the system's proper motion, $\mu = 4.5 \pm 0.8 \, \rm mas\,yr^{-1}$, which aligns with the value of $\mu = 4.97 \pm 0.02 \, \rm mas\,yr^{-1}$ measured from $Gaia$ Early Data Release 3 (EDR3) \citep{Vasiliev2021}.

The positive spin period derivative is discussed further in Section \ref{section:clustermodel}. Additionally, we have detected two derivatives of the orbital frequency, a characteristic commonly observed in BW and redback pulsar systems, not only in the Galactic disk \citep[e.g.][]{Shaifullah_2016} but also in GCs \citep[see e.g.][]{Ridolfi_2016,Rosenthal2024,Corcoran2024}. From these observations, we could also determine the orbital parameters of M92B, an MSP with $P=3.51 \,\rm ms$ in a 2.29-day mildly eccentric ($e \sim 10^{-4}$) orbit. Detailed analysis of this pulsar is under preparation (Yin et al. 2025).

\subsection{Properties of the Pulsar in NGC~6712}

NGC~6712A is a 2.15-ms pulsar in a BW system with an orbital period of 3.6 hr. As shown in Fig.~\ref{fig:xray}, this pulsar is well inside the core of the GC, a distance of only 0.16 core radii from the centre.
Due to the completely screening by the persistent luminous LMXB X~1850$-$086 (peak $L_{\rm X} \gtrsim 10^{35} \rm \, erg\,s^{-1}$; \citealt{Homer1996}) near the center of this cluster, the X-ray counterpart of NGC 6712A is not detectable and no significant spatial association with X-ray photons is observed.

After the observations until September 2020 reported by \cite{Yan2021}, we carried out 9 additional FAST observations from {April 2021 to November 2024}, from which we derive the updated timing solution for NGC 6712A presented in Table.~\ref{table:timing6712}. The postfit timing residuals obtained with this solution are presented in Fig.~\ref{fig:timingresiduals}. 
The solution now includes the proper motion of this system, $\mu=5.61 \pm 0.39 \, \rm mas\,yr^{-1}$, which is consistent with the value ($\mu=5.56 \pm 0.03 \, \rm mas\,yr^{-1}$) measured from $Gaia$ EDR3 \citep{Vasiliev2021}.
Similar to M92A, we have detected two derivatives of the orbital frequency.
In Section~\ref{section:clustermodel}, we will discuss in more detail the implications of the negative spin period derivative with an analytical cluster model that is not discussed in \cite{Yan2021}.

\begin{figure*}
\centering
	\includegraphics[width=18cm,height=11cm]{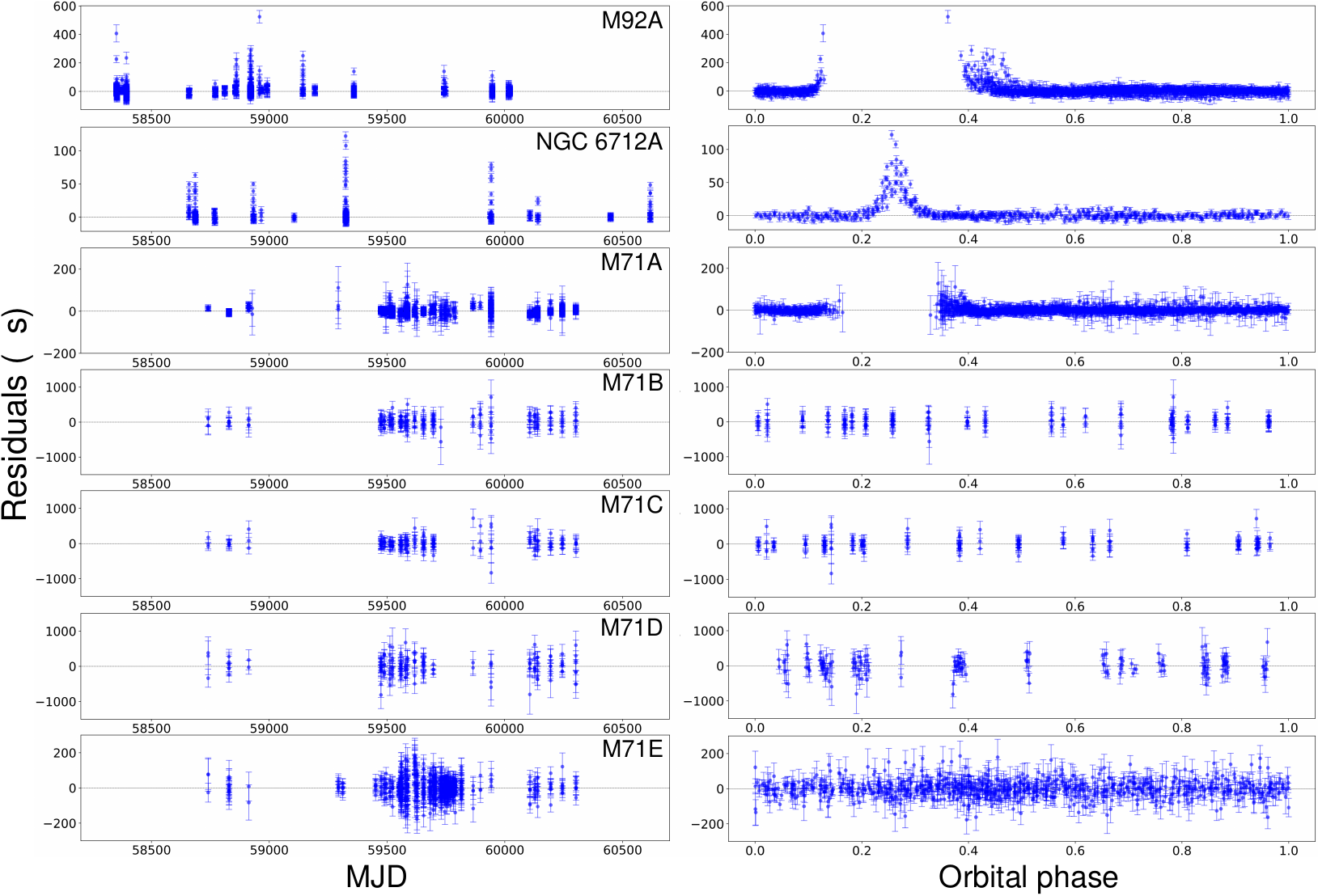}
	\caption{Timing residuals from the best-fit timing models presented in Table~\ref{table:timing6712} and Table~\ref{table:timing} as a function of the observation date (MJD, left panel) and orbital phase (right panel) for M92A, NGC 6712A, and M71A to E. Affected by the companion atmosphere, the TOAs of NGC 6712A around orbital phase 0.25 have an extra delay, which is very likely caused by ionized gas emanating from the companion \citep{Yan2021}. In the timing analysis, we exclude TOAs during eclipse phases: orbital phases 0.15$–$0.45 (M92A), 0.20$–$0.32 (NGC 6712A), and 0.15$–$0.34 (M71A). However, these excess delays remain visible in the residuals to illustrate the eclipsing signature. }
        \label{fig:timingresiduals}
\end{figure*}

\begin{figure*}
\centering
    \includegraphics[width=2.1\columnwidth]{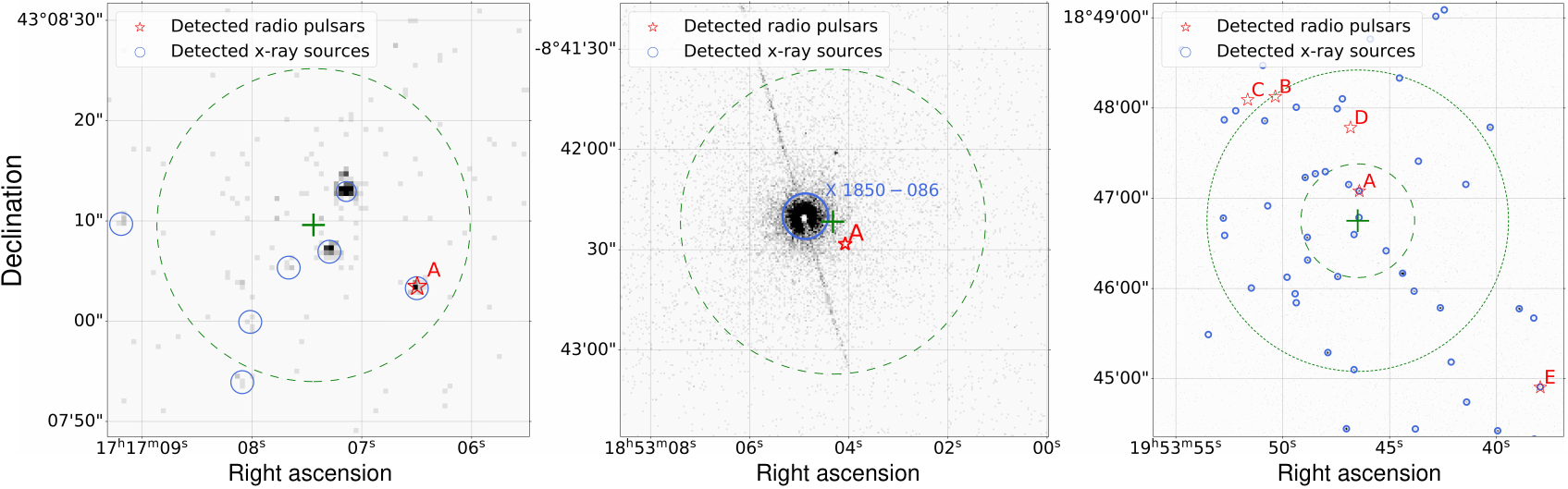}
    \caption{Positions of M92A, NGC 6712A, and M71A-E (denoted as red stars), and detected X-ray sources (denoted as blue circles), respectively. The details of those marked X-ray sources in M92, NGC 6712, and M71 can be found in \citet{Lu2011}, \citet{Geffert1994}, and \citet{Elsner2008} (and references therein), respectively. The center of these clusters (from \citealt{Harris2010}) is shown as a green cross. The core radii (0.26\arcmin, 0.76\arcmin, and 0.63\arcmin for M92, NGC 6712, and M71, respectively) are represented by green dashed circles. A half-light radius (1.67\arcmin) for M71 is denoted as a green dotted circle. The background images are X-ray observations from the \textit{Chandra} archive: M92 (OBsID 3778), NGC~6712 (OBsID 17849), and M71 (OBsID 5434). For reference, the beam size of FAST is 3\arcmin.}
    \label{fig:xray}
\end{figure*}

\subsection{Properties of Pulsars in M71}
\label{sec:M71}

With five known pulsars, M71 is the nearest GC and among the least dense clusters within FAST's sky (see Table~\ref{table:targets}). As seen in that table, it is now the least dense GC with known pulsars, with a central density significantly lower than even M53. As discussed below, this characteristic is reflected in the properties of its pulsar population.

M71A was discovered in a 1.4\,GHz Arecibo survey for GC pulsars \citep{Hessels2007}. It is a 4.89-ms BW pulsar in a 4.24-hr orbit, for which the timing solution and optical identification were later determined by \cite{Cadelano2015}. Its X-ray counterpart was identified by \citet{Elsner2008}.
M71B, C, and D have been reported in our former survey paper \citep{Pan2021b}, while M71E was discovered by \cite{Han2021} and confirmed in \citet{Pan2023}. 
A potential X-ray counterpart to M71E was discussed by \citet{Pan2023}. M71B is a 79.9\,ms binary in a mildly eccentric orbit and has the longest orbital period ($\sim$ 466\,days) among all known GC binaries, with a minimum companion mass ($M_{\rm c,min}$) of 0.42\,$\Msun$; assuming a pulsar mass 1.4\,$\Msun$ and inclination angle $i=90^{\circ}$. M71C is also in a low-eccentricity orbit with $P \sim 28.93~ \rm ms$ and the second longest orbital period $P_{\rm b} \sim 378 \, \rm days$, again with a low-mass companion ($ M_{\rm c,min}=0.36 \, \Msun$ with the assumption above). M71D is a 100.7\,ms pulsar in an eccentric orbit ($e \sim 0.63$), with an orbital period of $\sim 11$\,days and a
massive companion (more details below).

With 21 FAST observations from September 2019 to December 2023, we first confirmed M71D and obtained the phase-connected timing solutions for M71B, C, and D, and obtained refined timing results of M71A and E. All of these are presented in Table~\ref{table:timing}. In Fig.~\ref{fig:timingresiduals}, we display the post-fit timing residuals with all of the times of arrival (ToAs) obtained by these solutions. The positions of M71A-E are superposed on an X-ray map of the cluster in Fig.~\ref{fig:xray}. M71B, C, and D are located outside the core of M71, at distances $\gtrsim 1.6$ core radii from the center. With our best-fit positions with 1$\sigma$ uncertainty ($\sim10^{-2} \, \rm arcsec$ for R.A. and decl.), we found no associated X-ray counterparts of M71B-D. For M71D, we have a robust measurement of the rate of advance of periastron, $\dot{\omega}=0.0116(2)~\rm deg \; yr^{-1}$, which assuming the effect is relativistic results in a total system mass of $2.63 \pm 0.08 \, \Msun$. In Fig.~\ref{fig:DNSmap}, we present the 2D posterior probability distribution functions (pdfs) in the $\cos i$-$M_{\rm c}$ and $M_{\rm p}$-$M_{\rm c}$ planes, from which we also derived the 1D pdfs of a few quantities with their medians and 1$\sigma$ error. We get $M_{\rm p}=1.21_{-0.46}^{+0.13} \, \Msun$, $M_{\rm c}=1.41_{-0.13}^{+0.46} \, \Msun$. This indicates that M71D might be a DNS system. However, no other PK parameters are readily measurable. Our estimates indicate that, given the relatively low timing precision of this pulsar, no other relativistic effects will be measurable in the foreseeable future.

\begin{figure*}
\centering
	\includegraphics[width=0.8\textwidth]{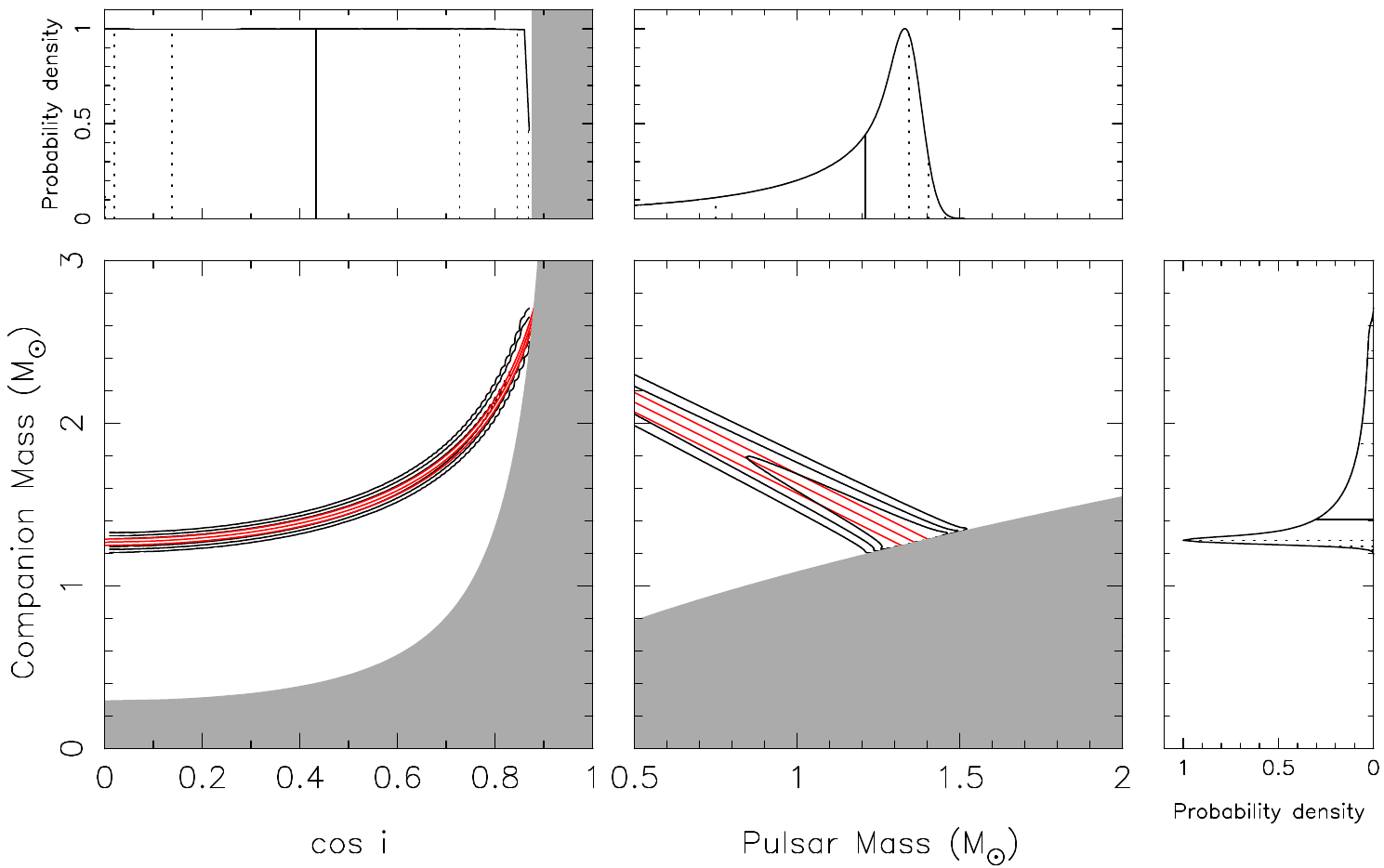}
	\caption{Mass-inclination and mass-mass diagrams for M71D. The contours display the $\cos i$--$M_{\rm c}$ and $M_{\rm p}$--$M_{\rm c}$ planes, including 1$\sigma$, 2$\sigma$, and 3$\sigma$ of a 2D pdf derived from the $\chi^2$ of $\tt TEMPO$ fit and the Bayesian technique described by \citet{Splaver2002,Freire2011b}. The red region shows the masses consistent with $\dot{\omega}=0.0116 \pm 0.0002 \, \rm deg \; yr^{-1}$, assuming GR is valid. In $\cos i$-$M_{\rm c}$ plane, the grey areas are excluded by the condition that the pulsar mass must be positive; in $M_{\rm p}$-$M_{\rm c}$ plane, they are excluded by the mass function and the condition $\rm sini \leq 1$. The 1D probability density functions for the $\cos i$ (top left), $M_{\rm p}$ (top right), and $M_{\rm c}$ (right) are obtained from the marginalized 2D pdf. The orbital inclination remains unconstrained, as the median of the $\cos i$ probability density function is close to 0.5. The timing result for M71D shows no significant Shapiro delay or any other relativistic effects that could allow the determination of individual masses.}
    \label{fig:DNSmap}
\end{figure*}

\begin{figure*}
\centering
    \includegraphics[width=2.1\columnwidth]{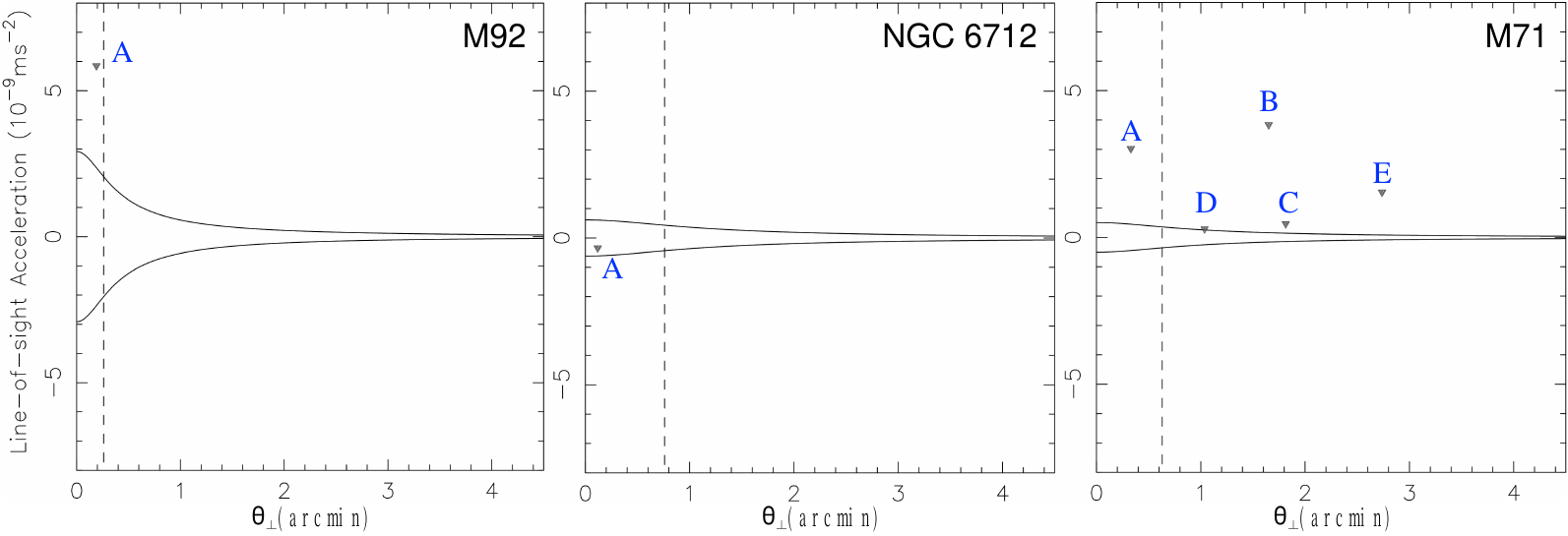}
    \caption{The acceleration model for M92, NGC 6712, and M71. The black solid lines represent the upper and lower limits for the line-of-sight accelerations ($a_{\ell \rm GC}$) caused by the cluster as a function of the total angular offset from the center of the cluster ($\theta_{\perp}$). The triangles pointing down represent independent upper limits for the pulsar accelerations along the line-of-sight, $a_{\ell, \rm P, max}$. The vertical dashed lines represent the core radii of the different GCs. The cluster models can only account for the acceleration of NGC 6712A and M71D. All the remaining pulsars have values of $a_{\ell, \rm P, max}$ well above the model's prediction, indicating that their observed spin period derivatives are dominated by the intrinsic spin-down of these pulsars.}
    \label{fig:acceleration}
\end{figure*}

\subsection{Pulsar Accelerations Caused by Cluster Potential}
\label{section:clustermodel}

In Tables~\ref{table:timing6712} and~\ref{table:timing}, we see that 
NGC~6712A has a negative $\dot{P}$, while for M92A and all pulsars in M71, this is positive.
Since the intrinsic spin period derivative $\dot{P}_{\rm int}$ is positive for radio pulsars, we want to investigate whether these pulsars are primarily accelerating due to the intrinsic spin-down or other gravitational fields.
The observed and intrinsic spin period derivatives ($\dot{P}_{\rm obs}$ and $\dot{P}_{\rm int}$) are related by:
\begin{equation}
\left( \frac{\dot{P}}{P} \right)_{\rm obs} = \left( \frac{\dot{P}}{P} \right)_{\rm int} + \frac{\mu^2d}{c} + \frac{a_{\ell,\, \rm GC}}{c} + \frac{a}{c},
\end{equation}
where $\mu$ is the total proper motion of the system, $d$ is the distance to the cluster ($\mu^2d / c$ is so-called the Shklovskii effect; see \citealt{Shklovskii1970}), $c$ is the speed of light, $a_{\ell,\, \rm GC}$ is the LOS acceleration of the pulsar in the gravitational field of the cluster, and $a$ is the acceleration of the center of mass of the GC in the potential of the Galaxy minus the Galactic acceleration of the solar system projected along the line-of-sight from the Earth to M92, NGC~6712, and M71.

Applying the Galactic mass model derived by \citet{McMillan2017}, we find accelerations due to the Galactic potential of $ a \, \sim \, -0.165 \times 10^{-9}$, $-0.149 \times 10^{-9}$, and $ -0.103 \times 10^{-9}\, \rm m \, s^{-2}$ for M92A, NGC~6712, and M71, respectively.
We have corrected the spin-down of M92A and NGC~6712A using their measured proper motion.
The proper motions of the pulsars in M71 have not yet been measured, but they are expected to be very similar to that of their cluster, $-3.416 \pm 0.025 \, \rm mas ~ yr^{-1}$ and $-2.656 \pm 0.024 \, \rm mas ~ yr^{-1}$ (from $Gaia$ EDR3; \citealt{Vasiliev2021}); thus $\mu^2d$ is $5.454 \times 10^{-11} \, \rm m \, s^{-2}$ was also taken into account in our calculations for the M71 pulsars.

To model the acceleration caused by the field of the GC, $a_{\ell, \rm GC}$, an analytical model of the cluster developed by \citet{Freire2005} is utilized, which assumes the \citet{King1962} density profile
\begin{equation}
a_{\rm GC}(x) = \frac{9 v_0^2}{d \theta_c} \frac{1}{x^2}
\left( \frac{x}{\sqrt{1+x^2}}  - \sinh^{-1}x \right),
\end{equation}
where $x$ denotes the distance from the pulsar to the center of the GC divided by its core radius $r_{\rm c}=\theta_c d$.
The acceleration along the line-of-sight, $a_{\ell, \rm GC}(x)$, can be computed by multiplying $a_{\rm GC}(x)$ with a factor of $\ell / x$, where $\ell$ is the separation along the line-of-sight direction between the pulsar and the center of gravity (COG) of the cluster, also in units of $r_{\rm c}$. For each line-of-sight, we calculate the maximum and minimum values of $a_{\ell, \rm GC}(x)$, $a_{\ell, \rm max}$. This type of model is valid for the inner regions of non-core collapsed GCs, and has been shown to provide a good description of the observed accelerations of pulsars near the cores of GCs \citep[see e.g., Fig. 6 of][which depicts the accelerations of pulsars in 47 Tucanae]{Freire2017}. This is also the case for all new timing solutions discussed in this work. In our models, we used the parameters of M92, NGC~6712, and M71 reported by \citet{Harris2010}, including the position (J2000 17:17:07.39, +43:08:09.4, 18:53:04.30, $-$08:42:22, and 19:53:46.49, +18:46:45.1 respectively), distance from the Sun ($d = 8.3$, 6.9, and 4.0\,kpc), their core radii ($\theta_c \, = \, 0.26\arcmin$, 0.76\arcmin, and 0.63\arcmin) and the central stellar velocity dispersion ($v_0 \, \sim \, 6.0$, $4.3$, and $2.3 \, \rm km \, s^{-1}$).

In Fig.~\ref{fig:acceleration}, the solid black lines represent the values of $a_{\ell, \rm max}$ as a function of angular distance from the center ($\theta_\perp$) of M92, NGC 6712, and M71. The values of $a_{\ell, \rm max}$ for the lines of sight of each pulsar are listed in Table~\ref{table:accelerations}. For all pulsars, an independent upper limit on the acceleration is determined from $\dot{P}_{\rm obs}$ as 
\begin{equation}
a_{\ell, \rm P, max} \, = \, c \frac{\dot{P}_{\rm obs}}{P} - \mu^2 d - a,
\end{equation}
which assumes $\dot{P}_\mathrm{int} = 0$. The values of $a_{\ell, \rm P, max}$ for each pulsar are also listed in Table~\ref{table:accelerations}. In Fig.~\ref{fig:acceleration}, they are represented as triangles pointing down, to emphasize that that they are upper limits. From this figure, we conclude that the range of accelerations predicted by the analytical models of M92 and M71 ($a_{\ell, \rm max}$) represent only a small part of the values of $a_{\ell, \rm P, max}$. This indicates that for M92A and the pulsars in M71 the positive $\dot{P}_{\rm int}$ is the dominant contribution to $\dot{P}_\mathrm{obs}$\footnote{Other contributions, like accelerations from nearby stars, are generally much smaller than the acceleration caused by the mean field of the GC \citep[e.g.,][]{Phinney1992}}. This is not the case for NGC~6712: although the values of $a_{\ell, \rm max}$ are not much larger than for M71, they are certainly larger than $a_{\ell, \rm P, max}$, because for NGC 6712A $\dot{P}_\mathrm{obs}$ and $a_{\ell, \rm P, max}$ are negative. Note that the gravitational model of NGC~6712 can account for the negative $a_{\ell, \rm P, max}$ of this pulsar.

By assuming for each pulsar that the accelerations are the maximum and minimum allowed by the mass model of the cluster for their lines of sight ($\pm a_{\ell, \rm max}$), we can calculate lower and upper limits for $\dot{P}_{\rm int}$, which in turn allow us to determine minimum and maximum values for their magnetic fields and maximum and minimum values for their characteristic ages. 
These results are detailed in Table~\ref{table:accelerations}. 
Unlike for other more massive GCs, the fact that $\dot{P}_{\rm int}$ is dominant for most pulsars means that we can estimate astrophysically useful limits.
From these limits, it is evident that all pulsars in M92, NGC~6712, and M71 have weak magnetic fields,
like similar systems observed in the Galactic disk; in particular, the magnetic field of NGC~6712A has an upper limit of only $6.4 \times 10^7$ G.
It is interesting that M71A, B, and E are consistent with a formation in the last 3 billion years, 
M92A has a characteristic age smaller than 1.36\,Gyr, i.e, less than 10\% of M92's age ($14.2 \pm 1.2 \,\rm Gyr$; \citealt{Paust2007}).
The larger characteristic ages of NGC~6712A, M71C and D imply that their real ages are unconstrained, in particular this means that we cannot exclude a formation of these systems at the early stages of the evolution of their clusters.

\begin{table*}[t]
    \centering
    \setlength{\tabcolsep}{4mm}{
        \begin{tabular}{cccccc} 
            \hline
            Pulsar name & $a_{\ell, \rm P, max}$ & $|a_{\ell, \rm max}|$ & $\dot{P}_{\rm int}$ & $B$ & $\tau_{\rm c}$ \\
                & $ 10^{-9} \rm m \, s^{-2}$ & $ 10^{-9}\, \rm  m \, s^{-2}$ &  $ 10^{-20} \, \rm s \, s^{-1}$ & $10^9$ G & Gyr  \\
            \hline  
            M92A  &   5.81  &  2.36   &  3.68 - 8.65  &  3.44 - 5.27   &  0.58 - 1.36  \\
       NGC 6712A  & $-$0.34 &  0.60   &  0 - 0.19     &  0    - 0.06   &  17.9 - $\infty$ \\
            M71A  &   2.97  &  0.33   &  4.20 - 5.66  &  0.46 - 0.53   &  1.37 - 1.85 \\
            M71B  &   3.78  &  0.11   &  98 - 106     &  8.93 - 9.30   &  1.19 - 1.29 \\
            M71C  &   0.41  &  0.10   &  3.10 - 5.84  &  0.96 - 1.31   &  7.86 - 14   \\
            M71D  &   0.23  &  0.18   &  1.22 - 18    &  1.12 - 4.31   &  8.80 - 131  \\
            M71E  &   1.48  &  0.06   &  2.15 - 2.40  &  0.31 - 0.33   &  2.93 - 3.27 \\
            \hline
        \end{tabular}}
    \caption{For each of pulsars in M71, we calculate the upper limit for the pulsar accelerations ($a_{\ell, \rm P, max}$), the absolute limits for the LOS pulsar acceleration due to the cluster potential ($|a_{\ell, \rm max}|$), and the lower and upper limits on the intrinsic spin period derivative ($\dot{P}_{\rm int}$), the surface magnetic field strength ($B$), and the characteristic age ($\tau_c$) respectively.
    \label{table:accelerations}}
\end{table*}

\subsection{Eccentricities of M71B, C, and D}

For orbital periods of the binary pulsars in GCs spanning a few days to a few hundred days, \citet{Phinney1992} predicts eccentricities between $\sim 10^{-6}$ to $\sim 10^{-3}$, yet the binaries in M71 display notably higher eccentricities $e \sim 10^{-4}-0.63$. Given the cluster's low stellar density, it is essential to assess if these eccentricities could result from the stellar interactions. Introduced by \citet{Rasio1995, Lynch2011}, the timescale required to produce the eccentricity of M71B, C, and D can be estimated as
\begin{eqnarray}
\nonumber t_{>e} &&\simeq 4 \times 10^{11}\; \yr 
\left (\frac{n}{10^4\; \pc^{-3}} \right )^{-1} 
\left (\frac{v_0}{10\; \km\, \ps} \right ) \\
&&\times\left (\frac{P\rmsub{b}}{\mathrm{days}} \right)^{-2/3}
e^{2/5},
\end{eqnarray}
where $n$ is the number density of stars ($n \propto \rho_c$, $\rho_c$ is the central mass density of the GC), 
and $P\rmsub{b}$ is the orbital period.
Assuming an average stellar mass of 1\,$\Msun$ \citep{Lynch2011}, the number density $n$ is roughly estimated as $n \approx 680 \, \rm  pc^{-3}$,
through the core density of M71 ($\rho_c \sim 0.68\times 10^3 \; \rm \Msun \, \rm pc^{-3}$; \citealt{Baumgardt2018}).
These values indicate $t_{>e} \approx 1.9$\,Gyr for M71B, $\approx 1.1$\,Gyr for M71C. These values roughly agree with the characteristic ages of M71B and C, and suggest that M71C could be younger than its characteristic age.
For M71D, we obtain $t_{>e} \approx 228$\,Gyr,  which is much larger than the age of the Galactic GCs ($\sim \rm 10 \, Gyr$), or the Universe, and indicates that the system formed with a large eccentricity that is very close to its current value.

\subsection{M71B and C: wide pulsar-He WD systems}

For the recycled, near-circular pulsar-He WD systems seen in the Galactic disk, the spin periods have some dependence on the orbital period. The longest orbital period for which we find spin periods below 10\,ms is
175 days, for PSR~J1640+2224, which has $P\, = \, 3.3$\,ms. Beyond this, all pulsars in low-eccentricity orbits with He WD companions have significantly slower spin periods and larger magnetic fields.
All systems are thought to have formed in Case B Roche lobe
overflow (RLO), where mass transfer happens after the companion leaves the main sequence, but before helium ignition \citep{Tauris_2023}; the difference is that for wider binaries, the time for which the companion
fills its Roche lobe is much smaller, resulting therefore in less spin-up and less ablation of the magnetic field.

In this respect, M71B and C are remarkably similar to the wide Galactic MSP - He WD systems: the minimum companion masses (0.42 and 0.36\,$\Msun$) are close to the prediction of the \cite{Tauris1999} relation for the masses of He WD companions in binaries with orbital periods of 466 and 378 days respectively. In particular, for M71B the close similarity between the minimum mass and the Tauris \& Savonije prediction indicates that the system is likely being observed with a high orbital inclination. The spin periods (79.9 and 28.9\,ms), magnetic fields, and even the orbital eccentricities are also very similar to the Galactic systems. The low eccentricities show that, despite their large age (of the order of Gyr), the perturbations by other stars in the GC have not raised the orbital eccentricities significantly, even for such wide binaries, which have a much larger cross-section for such interactions. 

None of this precludes the possibility that the MS star donor was acquired earlier in a dynamical interaction, as for most other pulsars in GCs: indeed the relatively small characteristic age of M71B ($\sim 0.1$ of the age of the GC) requires a relatively recent exchange encounter to form its progenitor LMXB. For a GC with such a low stellar interaction rate, such exchanges are not very likely for tighter binaries, which have small cross sections for such interactions, but will happen more often for wider, more weakly bound systems. This might be the reason why, despite its very low density, M71 has a relatively large number of pulsars, and why these wide binaries represent such a large fraction of all pulsars in M71. These two binaries are relatively rare cases in GCs: thus far the only known similar systems were the aforementioned M53A, a 33-ms pulsar in a 256-day orbit located in the low-density GC M53 \citep{Kulkarni1991} and B1620$-$26, a 11-ms pulsar in a 191-day orbit located in the low-density GC M4 \citep{Sigurdsson_2003}\footnote{The latter system likely has a more distant planetary-mass component}. The lower density of the core of M71 seems to correlate somehow with the larger percentage and orbital periods of these wide systems. Clearly, such wide binaries stand a much better chance of survival in these low-density GCs.

\subsection{M71D: a double neutron star system coeval with its host cluster?}
\label{sec:M71D}

The discovery of a DNS-like system in M71, a GC with such a low exchange encounter per binary, is unexpected. Until now, the candidate DNSs in GCs have been confined to GCs with a high stellar encounter rate per binary, where they can form dynamically. All such systems, like the aforementioned PSR~J0514$-$4002A and PSR~J1807$-$2500B, show abundant evidence of this: their spin periods and B-fields are much smaller than observed among Galactic DNSs, suggesting evolution in a normal MSP - WD system and latter disruption by a close encounter with a massive degenerate object, and the formation of an eccentric binary consisting of the MSP and the massive degenerate object.

M71D is very different. The low density of the GC makes it very unlikely it formed dynamically. Indeed, the large $t_{>e}$ calculated above shows that M71D was born with a very similar eccentricity to what it has nowadays, close encounters with other stars have changed its eccentricity very little.
Exchange encounters (which require even closer interactions) are therefore extremely unlikely. This suggests M71D was born in an eccentric system, and that this eccentricity was a result of its own unperturbed stellar evolution. The large orbital eccentricity, large companion mass, spin period, and magnetic field of M71D (somewhere between $1.1$ and $4.3 \, \times \, 10^{9}\, \rm G$) are very similar to those of the DNSs seen in the Galactic field. This suggests M71D is a DNS formed from the cluster's original stock of massive main-sequence star binaries in a way similar to the DNSs in the Galactic disk \citep{Tauris2017}.

There is another system in a GC that bears a remarkable resemblance to the DNSs in the Galactic disk, PSR~B2127+11C. However, it is safe to say that this system did not form like the DNSs in the Galactic disk because of its small characteristic age, which is less than 1\% the age of its host GC, M15 \citep{Prince1991}. No massive main sequence stars were available in M15 at that time to form this system. This means that, like other DNSs in GCs, M15C most likely formed dynamically. Importantly, this is not the case for M71D: its characteristic age is, depending on the unknown LOS acceleration of the system in the GC potential, somewhere between 8.8 and 131 Gyr. This is consistent with formation at the very early stages of the life of the M71 GC, when the massive stars necessary for the formation of such systems were still present. An additional line of evidence that this is a DNS system formed like the DNSs in the Galactic disk is its total mass. Indeed, the candidate DNS systems in GCs have a range of total masses that is much larger than the 
range of total masses seen among DNS systems in our Galaxy, the reason being that the companions might instead be massive white dwarfs \citep{Dutta_2025} or even possibly black holes (as in the case of PSR~J0514$-$4002E, see \citealt{Barr2024}). The total mass of the M71D system ($2.63 \pm 0.08 \, \Msun$) falls well within the range of total masses for the DNSs observed in the Galaxy.

\section{Discussion}
\label{sec:discussion}

In this section, we discuss the results of GC FANS, beginning with possible observational constraints (beam coverage, sensitivity) and their impact on detectability. We then link pulsar populations to cluster dynamics through stellar and single-binary encounter rates, analyze spin and binary period distributions in the context of recycling and dynamical interactions, and conclude with implications for the search for gamma-ray pulsations in M92, NGC 6712, and M71.

\subsection{Beam Coverage}

The spatial coverage of single-pointing observations can affect the detection of pulsars in GCs, particularly those located beyond 5 core radii ($r_{\rm c}$) from the cluster's center (see e.g., \citealt{Camilo2005}).
Here, we compare FAST’s beam coverage with that of MeerKAT to assess whether the single-pointing observations in GC FANS are sufficient to cover most visible pulsars in GCs. MeerKAT has a maximum spatial coverage of $\sim 5'$ with 288 coherent beams, which is enough to extend up to more than twice the reported half-light radius of a typical GC ($\sim 1'$) \citep{Abbate2022}.
This extensive coverage facilitates the detection of pulsars located far from the cluster center, such as PSR J1823$-$3022, found at $\sim 3'$ (nearly 50$r_{\rm c}$, where $r_{\rm c}=0.06'$), though it may not be associated with NGC 6624.
In contrast, we mainly used the central-beam observation (covering $\sim 3'$) at FAST in GC FANS. Despite this relatively small beam size, we detected NGC 6517H, I, and L at large distances from the center of this cluster ($r/r_{\rm c}\gtrsim 10$, with $r_{\rm c}=0.06'$) and M71E, located at $\sim 2.5'$ away from the center of M71 (about 4$r_{\rm c}$, where $r_{\rm c}=0.63'$).
Thus, the survey's spatial coverage is unlikely to have missed a significant number of pulsars due to FAST's effective beam size.

\subsection{Sensitivity to Weak Pulsars}

To evaluate FAST’s sensitivity to weak pulsars, we estimate the minimum detectable luminosity ($L_{\rm lim}$) of discovered pulsars in GC FANS, using the minimum flux density through Eq.~\ref{eq:fluxestimate}, $D_{\rm Sun}$ of GCs, the longest integration time of a single observation, and the best $S/N$ of pulsars.
Among the five clusters beyond 10\,kpc, the faintest MSPs detected have luminosities of $L_{\rm lim}=$ 0.23, 0.08, 0.11, 0.15, and 0.15\,mJy\,kpc$^2$ for M53D, M3E, NGC 6517Q, M15I, and M2D, respectively. These values place these pulsars among the least luminous known in distant clusters, providing an indication of the lowest luminosity limit for detectable pulsars in GCs.

In contrast, we also use $D_{\rm Sun}$ of GCs and the sensitivity estimation from Section~\ref{sec:sensitivity} to investigate the maximum luminosity ($L_{\rm max}$) of the brightest possible undiscovered pulsars in the surveyed clusters, assuming an isolated pulsar with $P=1 \, \rm ms$ (see Table~\ref{table:targets}).
For GC FANS clusters with known pulsars, $L_{\rm max}$ ranges from 0.03\,$\rm mJy \, kpc^2$ to 0.48\,$\rm mJy \, kpc^2$. These $L_{\rm max}$ values overlap with the faintest known pulsars in Terzan 5 (having 1.4\,GHz luminosity $L_{1400}\sim 0.7\,\rm mJy \, \kpc^2$; \citealt{Martsen2022}) and 47 Tucanae ($L_{1400}\sim 0.6\,\rm mJy \, \kpc^2$; \citealt{McConnell2004}). This indicate that those similarly faint pulsars should not be ruled out in our surveyed clusters, especially in those within $D_{\rm Sun}<18$\,kpc, such as M53.

\subsection{Stellar Encounter Rate}

Some studies have been devoted to forecasting the potentially observable GC pulsars \citep{Bagchi2011,Turk2013,Yin2024}, which correlate with various GC properties (e.g., cluster mass, metallicity, central density). 
Among these physical GC parameters, the stellar encounter rate $\Gamma$ (see Section.~\ref{sec:targets}) has emerged as a strong predictor of observable compact object populations. Observations indicate that $\Gamma$ correlates with the abundance of X-ray binaries \citep{Verbunt1987,Pooley2003} and radio pulsars \citep{Bahramian2013,Verbunt2014}, as dynamical interactions in high-$\Gamma$ clusters promote binary formation and NS recycling. Thus, high-$\Gamma$ GCs are expected to host more pulsars. However, this expectation had lead to more surveys being made in these clusters, which might have strengthened the correlation between the number of pulsars and $\Gamma$. To really test this expectation, unbiased GC surveys are extremely important.

In Fig.~\ref{fig:pulsarnum}, we map the Galactic distribution of 45 GCs within FAST's sky, highlighting the 14 clusters with known pulsars. Despite the established $\Gamma$-pulsar correlation, 60\% of our detected pulsars (36 out of 60) reside in lower-$\Gamma$ clusters ($\Gamma_{\rm M4}<10$). These findings imply that FAST's high sensitivity and comprehensive surveying enhanced the possibility of detecting pulsars in lower-$\Gamma$ clusters, where fewer pulsars were {\em a priori} expected (see e.g., \citealt{Bahramian2013}). However, the previous conclusion still stands: among high-$\Gamma$ clusters ($\Gamma_{\rm M4}>10$), we still see a clear positive trend between $\Gamma_{\rm M4}$ and total pulsar occurrence: NGC 6517 ($\Gamma_{\rm M4}=18.54$) hosts 21 pulsars (17 discovered in GC FANS), while M15 ($\Gamma_{\rm M4}=42.42$) hosts 15 pulsars (7 discovered in GC FANS). Although Pal 2 has the third-highest $\Gamma_{\rm M4}$ ($\sim 14.00$) in the FAST sky, its distance of 27.2\,kpc away from the Sun makes it difficult to find pulsars.

Unlike the simplified $\Gamma_{\rm M4}$ approach considered in this paper, \cite{Bahramian2013} directly derived stellar encounter rates by deprojecting observed GC surface brightness profiles, rather than scaling relative to M4's encounter rate. In their results, Terzan 5 (49 pulsars) and 47 Tuc (42 pulsars) have the highest $\Gamma$ values ($\Gamma=6800$ and $\Gamma=1000$, respectively), corresponding to the largest pulsar populations among the 45 GCs known to host pulsars. Within FAST's sky, NGC 6517 ranks third in $\Gamma$ value (338), trailing M15 ($\Gamma=4510$) and M2 ($\Gamma=518$), and hosts 21 pulsars, making it the third most pulsar-rich GC among the 45 GCs with detected pulsars.
Currently, no clusters with a stellar encounter rate of $\Gamma<13$ host a known pulsar, except for M71 (5 pulsars, $\Gamma=2.05$) and Terzan 1 (8 pulsars, $\Gamma=0.29$) \footnote{\url{https://www3.mpifr-bonn.mpg.de/staff/pfreire/GCpsr.html}}. This may be attributed to the relatively small distance of M71 and the high sensitivity of FAST ($D_{\rm sun}=4 \; \kpc$), as well as the dense core of Terzan 1 ($\rho_c \sim 7.1 \times 10^4 \, L_{\odot} \, \rm pc^{-3}$).

Among 41 GCs observed in GC FANS, it is not surprising that 16 clusters contain no pulsar, given their low $\Gamma$ value (0.001-9.4). Additionally, IC1257, Ko1, Ko2, and Whiting 1 have no estimation of $\Gamma$, as their core density have not been reported. Of the 14 clusters hosting pulsars, GLIMPSE-C01 does not have an available $\Gamma$ value; M10, M12, M13, M53, and NGC 6712 have a $\Gamma$ value between 13 to 69, accounting for $\sim$20\% of the total discoveries in this survey; M2, M3, M5, M14, M15, M92, and NGC 6517 exhibit $\Gamma$ values between 124 to 4510, accounting for $\sim$73\% of the total discoveries. Specifically, M2, M15, and NGC 6517, with the highest encounter rates and the greatest central escape velocities ($49–37\,\rm km\,s^{-1}$) in FAST's sky, were prioritized in this survey and are expected to host the largest pulsar populations in GC FANS \citep{Yin2024}.
Of the remaining 11 clusters in the FAST sky with interaction rates $\Gamma > 13$, where pulsars are expected to be found, none have had pulsars discovered. 
This lack of discoveries is likely due to their significant distances or sparse stellar density (e.g., $D_{\rm sun}\sim 76.5$\,kpc for Pal 14; $\rho_{\rm c }< 1 ~ \rm log \; L_{\odot} \rm \, pc^{-3}$ for NGC 5053). 

\begin{figure}
	\includegraphics[width=\columnwidth]{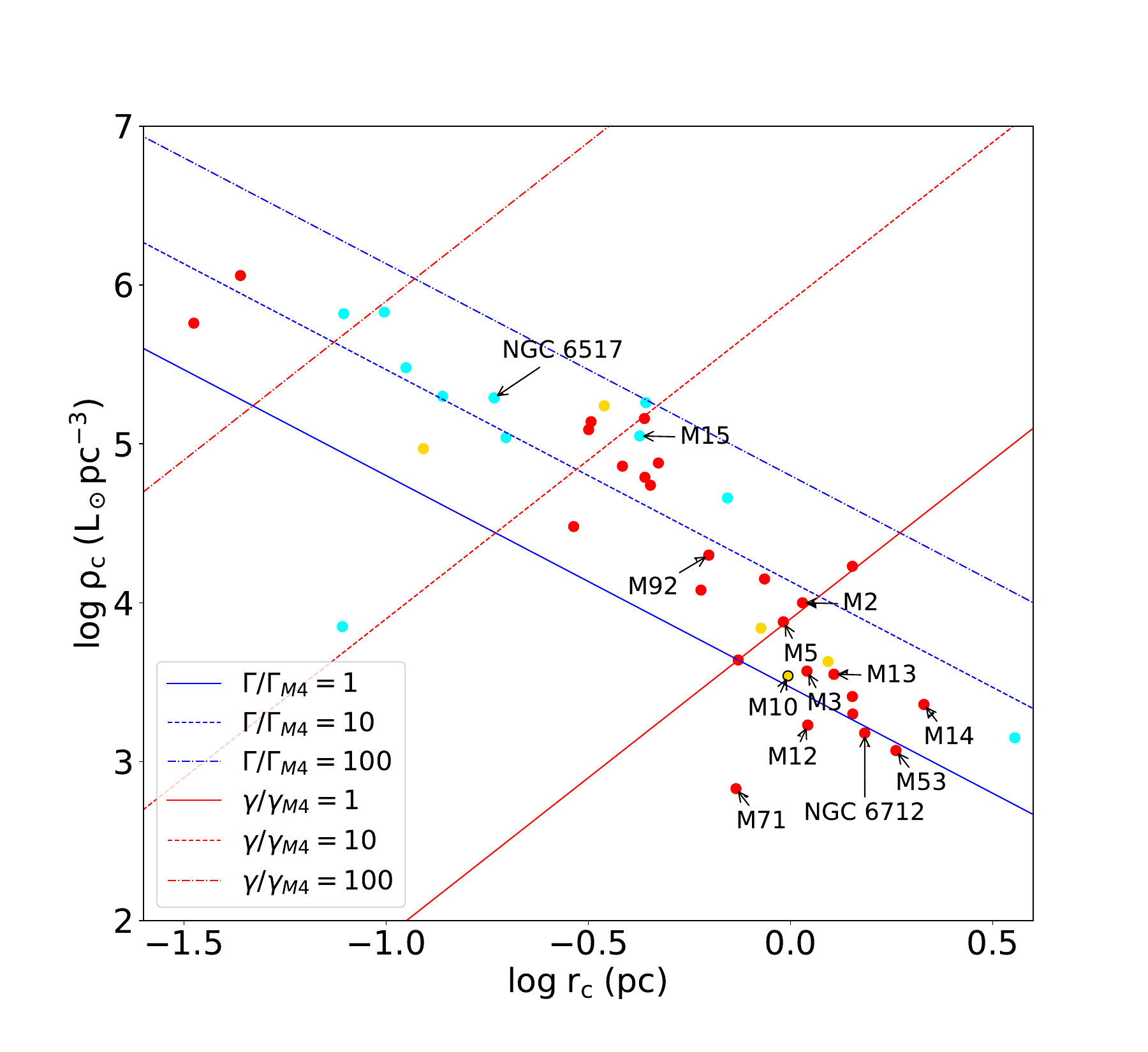}
	\caption{The central density as a function of the core radius for 44 clusters with known pulsars (excluding GLIMPSE-C01, which does not have available central density). The constant stellar interaction rate $\Gamma$ and encounter rate per binary $\gamma$ are shown as lines in the graph. Among these clusters, filled red, blue, and yellow circles indicate $N_{b}>N_{s}$, $N_{s}>N_{b}$, $N_{s} \simeq N_{b}$, where $N_{b}$ and $N_{s}$ are the number of binary pulsars and isolated pulsars respectively.}
        \label{fig:encounter}
\end{figure}

\subsection{The Single Binary Encounter Rate}
As discussed above, the stellar interaction rate for the whole cluster ($\Gamma$) and a single binary in GCs ($\gamma$) could let us better understand the numbers and population of radio pulsars respectively \citep{Verbunt2014}.
To evaluate the observed GC pulsar population in FAST sky, we discuss the single binary encounter rate for one object ($\gamma$), which refers to the secondary encounters in GCs, expressed as \citep{Verbunt2003}
\begin{equation}
\gamma \propto \frac{\rho_{\rm c}}{v_0} =C\frac{\sqrt{\rho_{\rm c}} }{r_{\rm c}},
\end{equation}
where $\rho_{\rm c}$ and $r_{\rm c}$ are the density and radius of the cluster core, and C is a constant. 

Fig.~\ref{fig:encounter} displays the central density as a function of core radius for 44 clusters with known pulsars.
The constant collision number ($\Gamma \propto \rho_{\rm c}^2 r_{\rm c}^3/v$) and the constant interaction rate per binary $\gamma$ are plotted as lines in a graph, scaling both to the values of the GC M4 as in \citet{Verbunt2014}.
Table~\ref{table:targets} lists the value of $\gamma$ for the clusters with FAST new discovery, most clusters have low $\gamma$ ($\rm \gamma \lesssim 1\gamma_{M4}$), including M2, M5, M10, M3, M13, M12, M71, NGC 6712, M14, and M53 (ordered on $\gamma$). Binary pulsars are found to dominate the pulsar population in these clusters with low-$\gamma$ (see Fig.~\ref{table:targets}). 
Only clusters M5, M13, and M53 have 1, 2, and 1 isolated pulsars found, which are all MSPs (see section 3). The high fraction of binary pulsars in these clusters is in good agreement with the prediction that low-$\gamma$ clusters resemble a similar MSP population in the Galactic disc, more specifically, the lower subsequent encounter rate makes the X-ray binaries evolve without much disruption. For instance, M53 has the second lowest single binary encounter rate $\rm \gamma=0.21\gamma_{M4}$ among all clusters with known pulsars, while the denser cluster Terzan 5 has a $\rm \gamma=13.0\gamma_{M4}$, which means there is a $\sim$ 60 times lower possibility of a close encounter for M53 compared to Terzan 5 per unit time. There are fewer than three known pulsars in M12, M10, and NGC 6712. This may bias the statistical discussion, and sensitive surveys are essential to finding more pulsars in such low-$\gamma$ clusters, which are predicted to have a smaller pulsar population than Terzan 5 (e.g., 5 pulsars are expected to be discovered in NGC 6712; \citealt{Yan2021}).

As for the intermediate to high $\gamma$ clusters, we found the FAST discoveries in M15 ($\rm \gamma=8.89 \gamma_{M4}$; 7 new isolated pulsars) and NGC 6517 ($\rm \gamma=26.82 \gamma_{M4}$; 17 new isolated pulsars) further highlight that their pulsar population are dominated by isolated pulsars, a consequence of the high rates of orbital disruption due to frequent encounters. This supports $\gamma$ (the encounter rate per binary) as a reasonable indicator of pulsar distribution in such core-collapse clusters. The frequent encounters not only disrupt binaries, but they might also disrupt LMXBs, leading to only a partial spin-up of the neutron stars. This would mimic the short recycling phase of high-mass X-ray binaries in the Galaxy (HMXBs), which result in spin periods of tens of ms or more that are seen in double neutron star systems in the Galactic disk. There are alternative possibilities for the formation of these objects: recent studies suggest that some isolated MSPs in GCs may form via a short-lived accretion episode following the tidal disruption of main-sequence stars, as derived from N-body simulations \citep{Kremer2022,Ye2024}, which again mimics the mild recycling produced by HMXBs.

However, as noticed by \citet{Verbunt2014} and discussed several times since (e.g., in $\omega$ Centauri by \citet{Chen2023} and the large populations of BW systems and isolated pulsars can not be fully characterized by $\Gamma$ and $\gamma$), extra factors should be complemented when characterizing the GC pulsar distribution, such as different evolution history of GCs. The surprising pulsar discoveries in low-$\Gamma$, low-$\gamma$ cluster M71 also indicate this. One factor might be that, irrespective of the $\Gamma$ of the cluster, wide binaries always have a larger cross-section to interactions than
tighter binaries, so for wider systems, the total interaction rates might still be reasonable. The low $\gamma$ means that binary pulsars formed in such wide systems (the examples being M71B and M71C) will more likely survive undisturbed for a long time (as indicated by the low eccentricities of these binaries), unlike in denser GCs, where such binaries will be much shorter-lived. The large number of wide binary pulsars implies that the number of pulsars in these GCs might not be as small as indicated by their $\Gamma$. In these clusters, we might even be seeing the preservation of coeval binary systems, like M71D, or the creation of compact systems like M71A and the extremely compact system M71E, such apparently exotic systems imply that even a very low $\gamma$ do not mean --- at least in some GCs --- the absence of highly eccentric or highly compact binary pulsars.

\begin{figure*}
\centering
	\includegraphics[width=18cm,height=8cm]{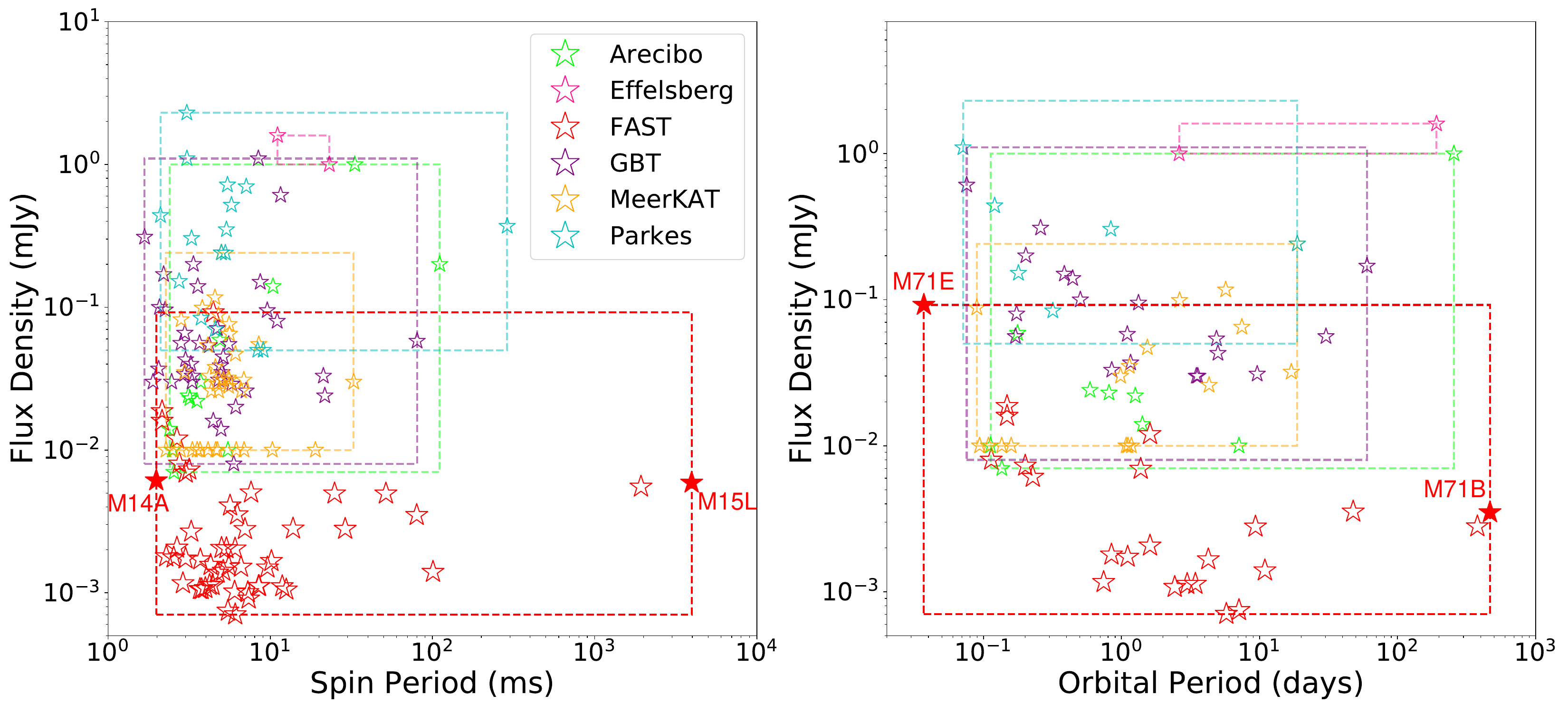}
	\caption{Flux density vs. pulsar spin period $P$ (left panel) and orbital period $P_{\rm b}$ (right panel) for pulsars in GCs discovered by different telescopes: Arecibo (lime), Effelsberg (deep pink), FAST (red), GBT (purple), MeerKAT (orange), and Parkes (cyan). The pulsars identified in the GC FANS demonstrate a notably deeper exploration of lower flux densities, with a wider range of $P$ and $P_{\rm b}$. The FAST observations highlight the minimum and maximum spin and orbital periods with solid markers. The dashed rectangles indicate the parameter space of spin and orbital periods vs. flux density of pulsars for each telescope. Most flux densities of pulsars at 1.4\,GHz come from the ATNF catalog, except for FAST, where flux densities are estimated using Eq.~\ref{eq:fluxestimate} and assumed 10\% of the pulse width, based on the best $S/N$, $P$, DM, and the longest integration time. Additionally, the estimated minimum flux densities of 13 pulsars in $\omega$ Centauri \citep{Chen2023} at 1.4\,GHz and 13 pulsars in NGC 1851 \citep{Ridolfi2022} at UHF are shown.}
        \label{fig:perioddistribution}
\end{figure*}

\subsection{The Distribution of Spin Periods and Binary Periods}

The left and right panels of Fig.~\ref{fig:perioddistribution} compare the spin and orbital period distributions for FAST-discovered GC pulsars with those from Arecibo, Effelsberg, GBT, and MeerKAT. It is clear that the vast majority of GC pulsars are MSPs, which is due to the frequent stellar interactions, the possibility of going through multiple stages of recycling, and the lack of recent star formation. Notably, GC pulsars discovered by FAST exhibit a broader spin period distribution ($1.98 \, {\rm ms} \lesssim P \lesssim 3.96 \, {\rm s}$) and lower flux densities compared to the general population. In GC FANS, we identified 10 rapidly spinning pulsars with $P \lesssim 3.0 \, \rm ms$.
Furthermore, although only a small number of slow pulsars ($P > 50~ \rm ms$) are known in GCs, we unexpectedly detected 5 such systems (M15K, M15L, M71B, M71D, and NGC 6517G), which account for about 25\% of all known slow GC pulsars.
Some studies indicate that such slow systems are generally found in higher-density clusters \citep{Johnston1991,Verbunt2014}, yet their origins remain unclear.
These long-period pulsars are apparently ``young" with a typical lifetime $\sim 10^7-10^8 \, \rm yr$ (e.g., J1823$-$3021C has a characteristic age of $\sim 28\, \rm Myr$; \citealt{Lynch2012}), which is in contradiction with the fact that their progenitor (i.e., core-collapse supernovae of massive stars) can not exist in GCs for billions of years. Aside from the disruption of X-ray binaries \citep{Verbunt2014}, other possible explanations for the formation of these slow GC pulsars include accretion-induced collapse \citep{Tauris2013}, direct collisions with a main sequence star \citep{Lyne1996}, electron capture supernova of an OMgNe white dwarf \citep{Boyles2011}, and white dwarf mergers \citep{Kremer2023}. As shown in the left panel of Fig.~\ref{fig:perioddistribution}, M15K and L are unusually slow with the longest spinning periods ($\sim 1.9, 3.9 \, \rm s$, respectively), ranking third and first among all GC pulsars. The discoveries of such slow pulsars in M15 could be a real effect related to the stellar interaction, instead of a selection effect \citep{Verbunt2014}.

The orbital periods of FAST discoveries spans from 0.037\,days (M71E; confirmed in GC FANS; \citealt{Pan2023}) to 466\,days (M71B), which is also broader than the general distribution as shown in Fig.~\ref{fig:perioddistribution}. Among 35 binaries first discovered in GC FANS (20 with measured orbital parameters), four (M2G, M14A, M92A, and NCG 6712A) are found in very tight orbits ($P_{\rm b} \leq 6 \, \rm hr$). All of the FAST discoveries always indicate a smaller flux density, which broadens the parameter space in the flux density-orbital period plane. Again, the discoveries of M71E and M71B highlight FAST's ability to probe extreme ends of pulsar characteristics, uncovering unique systems that may otherwise remain undetected. These findings provide critical insights into the diversity and evolution of binary pulsars in dense stellar environments. Still, in order to create such a large number of binary pulsars, M53 and M71 must have retained a significant number of NSs for subsequent interactions, which
suggests they were much more massive in the past - another factor that is not taken into account by their current values of $\Gamma$ and $\gamma$. This is even suggested by the retention of a DNS that is likely coeval with M71: indeed, these systems tend to form with large recoils \citep{Tauris2017}, so their retention in the GC should require a relatively large escape velocity, which M71 does not have at present (the central escape velocity is $\sim 10 \rm \, km \, s^{-1}$)\footnote{\url{https://people.smp.uq.edu.au/HolgerBaumgardt/globular/parameter.html}.}.

\begin{figure*}\
\centering
	\includegraphics[width=0.72\columnwidth]{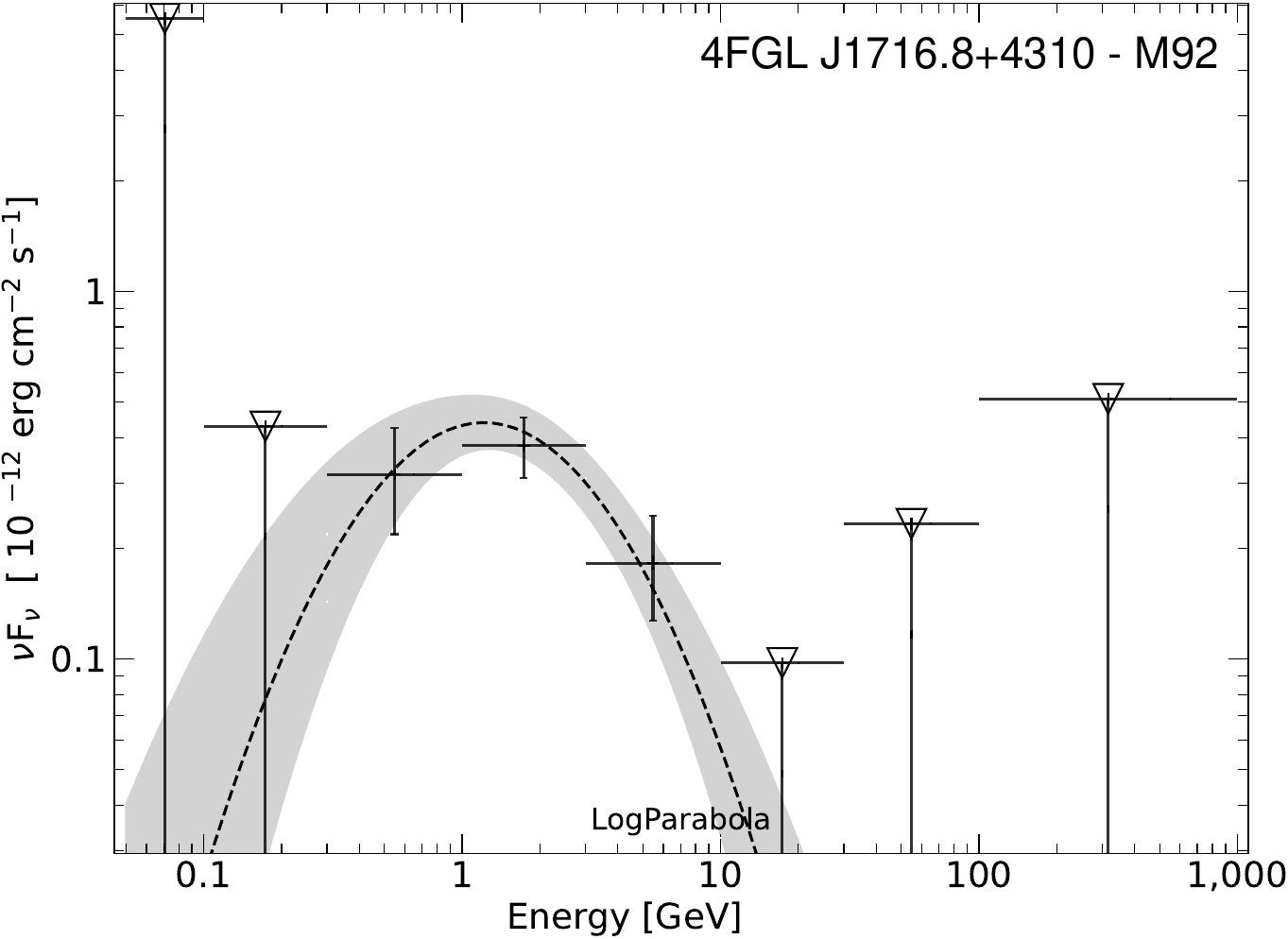}\includegraphics[width=0.72\columnwidth]{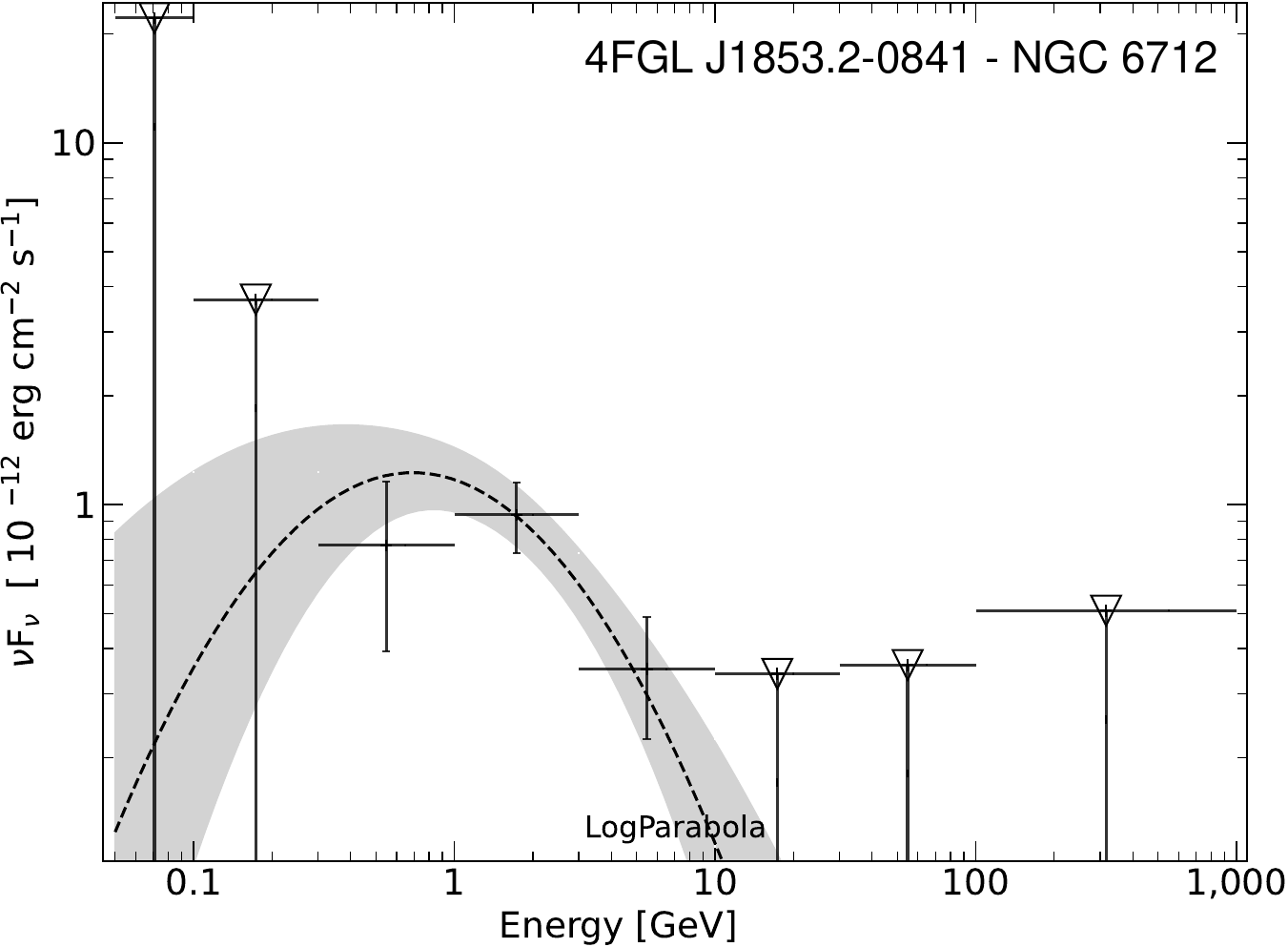}\includegraphics[width=0.72\columnwidth]{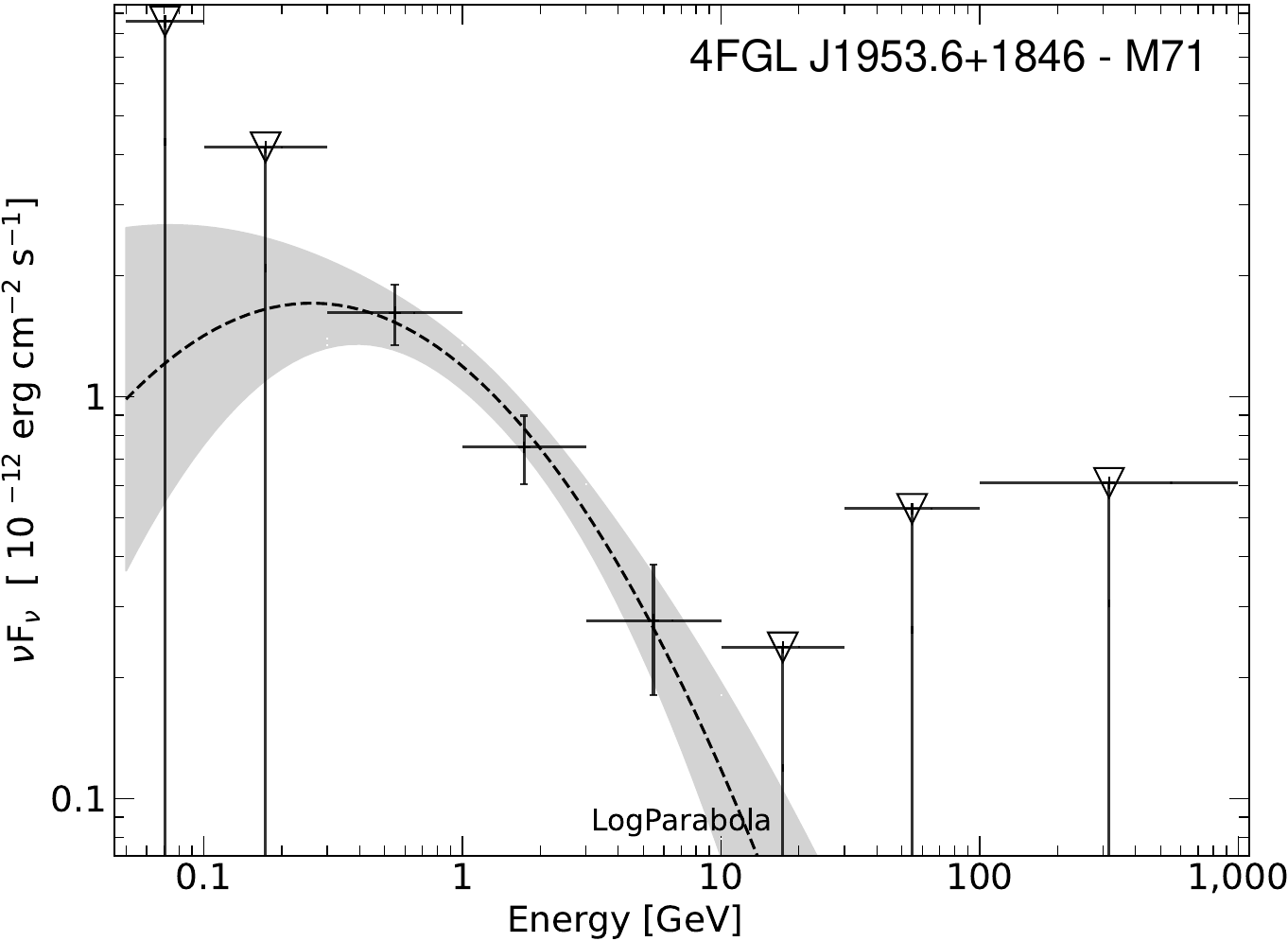}
	\caption{Gamma-ray spectral energy distributions (SEDs) for M92, NGC 6712, and M71, which are typical of SEDs observed for MSPs. The log-parabola fit to the data comes from the 4FGL-DR4 catalog \citep{Ballet2023}. The grey zone shows the 1\,$\sigma$ uncertainty range.}
    \label{fig:Fermi_SED}
\end{figure*}

\subsection{Gamma-rays from Globular Clusters}
\label{sec:gammas}

Nine of the GCs targeted by GC FANS are associated with GeV gamma-ray sources listed in the 4FGL-DR4 catalog \citep{Abdollahi2022,Ballet2023}, with available SED plots~\footnote{\url{https://fermi.gsfc.nasa.gov/ssc/data/access/lat/14yr_catalog/}}. They are GLIMPSE-C01, and NGC catalog numbers NGC 5904, 6205, 6218, 6341, 6402, 6712, 6838, and 7078 (see examples in Fig \ref{fig:Fermi_SED}).
Some studies have found $\gamma$-ray pulsations from individual MSPs in NGC 6624 \citep{Freire2011a}, NGC 6626 \citep{Johnson2013}, and NGC 6652 \citep{Zhang2022}.
\citet{Hou2024} recently searched for NGC 5904 (M5), but detected no significant gamma-ray pulsations from any individual pulsars.

The Fermi LAT pulsar catalog \citep[3PC]{Smith2023} includes the three MSPs in GCs just cited, outside of GC FANS's declination range. The spectral shape in Fig.~\ref{fig:Fermi_SED} is typical of gamma-ray pulsars and also of gamma-ray emitting GCs, see the discussion in Section 6 of 3PC.
3PC further shows that all known gamma-ray pulsars have spin-down energy loss rate $\dot{E} \gtrsim 10^{33}$ erg s$^{-1}$, with the fraction detected in gamma rays increasing with $\dot{E}$. A number of the GC FANS pulsars have large $\dot{E}$ and merit gamma-ray pulsation searches. 

For M92A, $\dot{E}$ is $0.46$ to $1.08 \times 10^{35} \, \rm erg\,s^{-1}$ (derived from $\dot{P}_{\rm int}$ values in Table~\ref{table:accelerations}). 
Using the earlier ephemeris of M92A from \cite{Pan2020}, \cite{Zhang2023b} reported a tentative detection of $\gamma$-ray pulsations from M92A.  
According to the non-detection of M92A's $\gamma$-ray pulsations in \citet{Smith2023}, we re-folded all Fermi-LAT data to date using our significantly extended ephemeris. No pulsed signal was found. This non-detection is likely due to M92A’s $\gamma$-ray emission beam not intersecting Earth or the total $\gamma$-ray flux being below Fermi-LAT's sensitivity threshold. Including M92A, we tried to fold the Fermi-LAT data with the timing of NGC 6712A, and M71A to E and found none of them either having a $H_{\rm max}>10$ (defined as the maximum H-test value for six trials; see details in \citealt{Smith2019}), or a significant $\gamma$-ray pulsations. Based on the $P$ and $\dot{P}$ values listed in Table~\ref{GC pulsars}, we roughly evaluated the $\dot{E}$ of the pulsars with a positive observed $\dot{P}$. M2D, M3F, M5G, M13F, M14A, M53C and D, and NGC 6517H show $\dot{E} > 10^{33}\, \rm erg\,s^{-1}$. While these pulsars are potential $\gamma$-ray candidates due to their relatively high $\dot{E}$, the gravitational potential of their host clusters could influence their intrinsic $\dot{P}$, potentially lowering their actual $\dot{E}$ below this threshold.

\section{Conclusions and prospects}
\label{sec:conclusions}

\subsection{Conclusions}

In this paper, we present the details of the seven-year GC FANS survey, including the source list, as well as the survey sensitivity, and summarize the 60 discoveries in 14 clusters within FAST's sky from 1.05\,GHz-1.45\,GHz. 
This more than doubles the number of known GC pulsars in the FAST sky (from 35 to 95). Owing to the high sensitivity of FAST, this survey is a deep, thorough search for GC pulsars in the northern hemisphere, including clusters with low stellar encounter rates. We have also reported the first phase-connected timing solutions of M71B to D, and updated the timing solutions of NGC 6712A, M92A, M71A, and E. This thorough strategy gives us an unprecedented view of the pulsar populations in low-density GCs, which is important for drawing general conclusions about the pulsar populations in GCs. It has also has been producing surprising results: An an example, despite the large distance of M53 - the largest for any GC with known pulsars ($d \, = \, 17.9\,$kpc) - and its very low stellar density - the second lowest for any GC with known pulsars - we have discovered 4 MSPs in this cluster.

Another prime example of the surprises provided by low-density GCs is the remarkable pulsar population in M71. This GC, with the lowest density of any GC with known pulsars (see Fig.~\ref{fig:encounter}), has at the same time, the binary pulsar with the smallest orbital period known (M71E, 0.037\,days), but also the two binary pulsars with the largest orbital periods known in GCs (M71B and C, 466 and 378\,days respectively). The mildly eccentric orbits of the latter two binaries render them potential targets for gravity tests. This population even includes M71D: with an eccentric orbit ($e \sim 0.63$), extremely large characteristic age and a total system mass of $2.63 \pm 0.08 \, \Msun$, this DNS likely formed via regular star formation channels in the very early stages of the evolution of this GC, not dynamically as for other candidate DNSs in GCs. This is the first such known case in a GC.

The cases of M53 and M71 confirm that some low-density GCs have more pulsars and more eccentric/compact systems than might be expected from their stellar encounter rates \cite[this was already noticed after the first Arecibo discoveries in M5, M13 and M53 by][]{Kulkarni1991}. Part of the reason must be the sensitivity of FAST: as an example of this, M53D is one of the weakest known radio pulsars, with a measured mean flux density of $\sim$ 0.7\,$\mu \rm Jy$, calculating through Eq.~\ref{eq:fluxestimate} with a 5-hr integration, its detected S/N$\sim$6, DM$\sim24.61 \, \dmu$, and $P\sim 6.07 \, \rm ms$.
In the case of M71 (but certainly not in the case of M53) the relative proximity also plays a role.
An additional reason might be the preservation of wide binary pulsars: they can form from wide LMXBs, which can form dynamically (via exchange or tidal capture by giant stars) with high probability even in these low-density clusters, but only in these low-density clusters are they likely to be preserved.
These surprising findings emphasize the benefits of thorough surveys for a full understanding of the pulsar population in GCs. Other reasons might have to do with the peculiar past histories of these GCs.

The discoveries of the GC FANS survey and its thorough nature have also strengthened some of the previously observed trends in the pulsar populations of GCs. Out of the 14 clusters with pulsar discoveries, 10 clusters have an encounter rate per binary $\gamma_{\rm M4} \lesssim 1$, in these binary pulsars are found to dominate the pulsar distribution. This is consistent with the expectation that the X-ray binaries in such clusters will evolve without much disturbance, forming binary MSPs similar to those of the Galactic disk. For the GCs with a much larger $\gamma_{\rm M4}$, especially for core-collapse clusters (such as NGC 6517 and M15 in this survey), isolated pulsars are observed to vastly dominate the pulsar distribution: out of a total population of 36 pulsars in these two GCs, only two binaries (one in each cluster) have been found. This confirms the predominance of isolated pulsars in core-collapsed GCs discussed by \citet{Verbunt2014}. This might be a consequence of the high rates of orbital disruption, caused by the dense surroundings. The spin periods of our discoveries, ranging from 1.98 ms to 3.96 s, expand the known distribution, particularly with the detection of two slow pulsars in M15 ($P > 1\,\rm s$). This indicates that such pulsars are generally found in higher-density clusters, again as noted by \cite{Verbunt2014}. Their formation and evolution have not yet been fully clarified, but they could be due to the disruption of LMXBs in the early stages of the accretion process, leaving behind mildly recycled pulsars.

\subsection{Gravity Tests from M71B and C}

Long-orbit and small-eccentricity binary pulsars are useful for tests of the strong equivalence principle~\citep[SEP; ][]{ds91} and the local Lorentz invariance (LLI) of the gravity sector~\citep{de92}, both of which are pillar elements of Einstein's general relativity~\citep{will18}. M71B and C could be potential interesting targets for these tests after a careful reconsideration of traditional methodologies by taking into account of the specifics of M71, augmented with a long-term, high-precision observation in the future.

In the case of SEP violation, the inertial mass and gravitational mass might differ for the strongly self-gravitating NS, which leads to a ``polarized'' binary, meaning that the observed orbital eccentricity vector is resulted from a vectorial superposition of a constant SEP-violating vector (characterized by the SEP-violating parameter $\Delta$) and a general-relativistically rotating one~\citep{ds91}. Such a scenario can be constrained by using long-orbit and small-eccentricity binary pulsar systems, either using a statistical approach or a direct test~\citep{fkw12}. Currently, the best direct test from binary pulsars uses 21-yr timing data of PSR~J1713+0747~\citep[$P_{\rm b} = 67.8\,$days and $e=0.000075$; ][]{zdw+19}. In terms of the figure of merit for the statistical SEP test, which is $\sim P_{\rm b}^2 / e$, M71B and C both surpass that of PSR~J1713+0747 by factors of 1.8 and 6.2, respectively. Therefore, they could become potential targets of the SEP test if in the future better timing results are obtained from a longer observational span. However, because of the movements of M71B and C in the GC, whose revolving periods are much shorter compared to those of typical binaries in the Galactic field, the original method by \citet{ds91} should be carefully reconsidered and rectified~\citep{wex00}. Furthermore, it was demonstrated that such kin SEP tests from binary pulsars could also be re-interpreted to put interesting limits on the long-range fifth force from dark matter, unreachable with terrestrial experiments~\citep{swk18}. Here in M71, the dark-matter environment is quite distinct from that of the Galaxy, and, depending on the detailed, yet unknown, nature of dark matters and their evolution, M71B and C could have potential to provide new, if not complementary, tests to what was done with PSR~J1713+0747 in the Galactic field. These tests of dark matter properties are not feasible with the triple pulsar system PSR~J0337+1715, which though has provided a much better SEP test than any other binary pulsars~\citep{agh+18,vcf+20}.

In the case of LLI violation, a similar ``gravitational Stark effect'' happens and it results in a polarized orbit, now the deviation being characterized by a strong-field counterpart of the parameterized post-Newtonian parameter, $\hat\alpha_1$~\citep{de92,sw12}. Again there are two approaches developed to constrain the $\hat\alpha_1$ parameter with binary pulsar timing, a statistical one and a direct one~\citep[see ][for more details]{sw12}. M71B and C have no compelling strength in the direct test but might have the possibility to provide a meaningful limit on $\hat\alpha_1$ in the statistical test, whose figure of merit is $\sim P_{\rm b}^{1/3} / e$. Different from the SEP test in the preceding paragraph, the $\hat\alpha_1$ test does not depend on the revolving movements of M71B and C in the GC if the Lorentz symmetry violating is cosmological or the associated length-scale is larger than the Galaxy. It is also interesting if, due to the detailed LLI-violating mechanism, the violation is localized to M71, similar to the case of a violation in the local rest frame of the Solar System, as explored in detail by \citet{sw12}. In these cases, M71B and C could be able to provide new aspects of modified gravity theories. Nevertheless, the current timing precision is quite preliminary, and future long-term observations are needed to better characterize the orbits.

Without diving into detailed simulations, timing parameters, $e$ and $\dot e$, improve with the observing time span $T_{\rm obs}$ as $T_{\rm obs}^{1/2}$ and $T_{\rm obs}^{3/2}$, respectively, if current timing precision and cadence are assumed \citep{dt92}. Considering the current uncertainty of $e$, $\sigma_e \sim {\cal O} \big( 10^{-7} \big)$ (see Table~\ref{table:timing}), as well as the likely improvements in the timing precision for the future, one might be able to measure the time derivatives of the EPS1 and EPS2 parameters, enabling direct tests of SEP and LLI violations discussed above. A quantitative projection of improvements and measurability needs to relax the timing model used for Table~\ref{table:timing} by including the time derivatives of the EPS1 and EPS2 parameters (see e.g.~\citealt{Liu2020} for PSR~J1909$-$3744) as well as providing a future observing cadence, which goes beyond the scope of the current work.

\subsection{Prospects}

We are conducting further observations to derive additional timing solutions for the pulsars found by GC FANS and refine those that have already been obtained. These will result in an improved understanding and multi-wavelength follow-up of these pulsars (e.g., searches from the Large Area Telescope on the Fermi Gamma-ray Space Telescope; \citealt{Smith2023}). Additional timing solutions will also be important for probing the gravitational potentials of these GCs, might provide new clues on their evolution, and promote testing alternative gravity theories as mentioned above.
As illustrated in Section~\ref{sec:data reduction}, pulsar search methodologies exhibit sensitivity to different types of pulsars (e.g., as discussed several times, jerk search is useful to catch the compact and ultra-compact binaries; \citealt{Andersen2018}). Future work will systematically reanalyze archival GC FANS observations to ensure an unbiased pulsar search (e.g., performing a jerk search across more archival observations for compact binaries). Finally, the use of more efficient search techniques (e.g., the 3D-template-bank-searched; \citealt{Balakrishnan2022}) will likely lead to the discovery of new, and possibly even more extreme binary pulsars in the FAST observations.

\section*{Acknowledgements} 
We sincerely appreciate the referee’s very insightful review and the many constructive suggestions, which have greatly improved the quality of this work. We thank David A. Smith for his valuable discussions on the Fermi-LAT search for pulsars found in GC FANS and for his assistance in re-folding the 4FGL-DR3 data using the ephemeris of timing updated in this study. This work is supported by National Key R\&D Program of China, No. 2022YFC2205202; National Natural Science Foundation of China under Grants Nos. 12021003, 12433001, 12041301, 12173052, 12173053, and 11920101003; Beijing Natural Science Foundation No. 1242021; the Fundamental Research Funds for the Central Universities; Guizhou Provincial Science and Technology Projects (No.QKHFQ[2023]003, No.QKHFQ[2024]001, No.QKHPTRC-ZDSYS[2023]003). Z. Pan and L. Qian are supported by the CAS ``Light of West China" Program and the Youth Innovation Promotion Association of the Chinese Academy of Sciences (ID 2023064, 2018075, and Y2022027), National SKA Program of China No. 2020SKA0120100. P.C.C.F. gratefully acknowledges continued support from the Max-Planck-Gesellschaft. R.P.E. is supported by the Chinese Academy of Sciences President's International Fellowship Initiative, grant No. 2021FSM0004. L.S. was supported by the National SKA Program of China (2020SKA0120300) and the Max Planck Partner Group Program funded by the Max Planck Society. Y. Wang and Z. Zhang are supported by National SKA Program of China No. 2020SKA0120200. L. Zhang was supported by the National Natural Science Foundation of China with No. 12373032, and the Guizhou Provincial Science and Technology Plan Project (No. ZDSYS[2023]003 and QKHFQ[2023]003). Y. Lian is supported by the China Scholarship Council (Grant No. 202306040148). This work made use of the data from FAST (\url{ https://cstr.cn/31116.02.FAST}). FAST is a Chinese national mega-science facility, operated by National Astronomical Observatories, Chinese Academy of Sciences. We appreciate all the assistance from FAST staffs during the observation. The National Radio Astronomy Observatory is a facility of the National Science Foundation operated under cooperative agreement by Associated Universities, Inc. SMR is a CIFAR Fellow and is supported by the NSF Physics Frontiers Center award 2020265.

\end{document}